\documentclass[a4paper,onecolumn,11pt,accepted=2021-07-21]{quantumarticle}
\pdfoutput=1
\usepackage[utf8]{inputenc}
\usepackage[english]{babel}
\usepackage[T1]{fontenc}
\usepackage{amsmath}
\usepackage{hyperref}
\usepackage[numbers,sort&compress]{natbib}

\usepackage{tikz}
\usepackage{lipsum}

\usepackage{graphicx}% Include figure files
\usepackage{dcolumn}% Align table columns on decimal point
\usepackage{bm}% bold math
\usepackage{hyperref}% add hypertext capabilities
\usepackage{mathtools}
\usepackage{color}
\usepackage{enumitem}
\usepackage{chemformula}
\usepackage{amssymb,amsmath,amsthm,amsfonts}
\usepackage[version=4]{mhchem}
\usepackage{chemfig}
\usepackage{expl3}
\input{Qcircuit}
\usepackage{multirow}
\usepackage{subfloat}
\usepackage{float}
\usepackage[normalem]{ulem}

\begin{document}

\newcommand{\eq}[1]{(\ref{eq:#1})}
\renewcommand{\sec}[1]{\hyperref[sec:#1]{Section~\ref*{sec:#1}}}
\newcommand{\fig}[1]{\hyperref[fig:#1]{Fig.~\ref*{fig:#1}}}
\newcommand{\tab}[1]{\hyperref[tab:#1]{Table~\ref*{tab:#1}}}
\newcommand{\routine}[1]{\hyperref[#1]{Routine~\ref*{#1}}}

\newcommand{\Gate}[1]{\textsc{#1}}
\newcommand{\hgate}{\Gate{h}}
\newcommand{\zgate}{\Gate{z}}
\newcommand{\sgate}{\Gate{s}}
\newcommand{\tgate}{\Gate{t}}
\newcommand{\pgate}{\Gate{p}}
\newcommand{\notgate}{\Gate{not}}
\newcommand{\cnotgate}{\Gate{cnot}}
\newcommand{\cnotgates}{\Gate{cnot}s}
\newcommand{\rzgate}{{\Gate{r}}_{z}}
\newcommand{\rygate}{{\Gate{r}}_{y}}
\newcommand{\rxgate}{{\Gate{r}}_{x}}
\newcommand{\gmsgate}{{\Gate{gms}}}
\newcommand{\czgate}{{\Gate{c-z}}}

\newcommand{\vph}{\vphantom{\Psi_{D_{\{}}}}

\newcommand*{\Perm}[2]{{}^{#1}\!P_{#2}}%
\newcommand*{\Comb}[2]{{}^{#1}C_{#2}}%
\newcommand\sbullet[1][.5]{\mathbin{\vcenter{\hbox{\scalebox{#1}{$\bullet$}}}}}

\title{Resource-Optimized Fermionic Local-Hamiltonian Simulation on a Quantum Computer for Quantum Chemistry}

% Author 1
\author{Qingfeng Wang}
\email{keewang@umd.edu}
\affiliation{Chemical Physics Program and Institute for Physical Science and Technology, University of Maryland, College Park, MD 20742, USA}

% Author 2
\author{Ming Li}
\email{li@ionq.co}
\affiliation{IonQ, College Park, MD 20740, USA}
\orcid{0000-0003-0827-5976}

% Author 3
\author{Christopher Monroe}
\email{monroe@umd.edu}
\affiliation{IonQ, College Park, MD 20740, USA}

\affiliation{Joint Quantum Institute, Department of Physics, and Joint Center for Quantum Information and Computer Science,
University of Maryland, College Park, MD 20742, USA}

% Author 4
\author{Yunseong Nam}
\email{nam@ionq.co}
\affiliation{IonQ, College Park, MD 20740, USA}
\affiliation{Department of Physics,
University of Maryland, College Park, MD 20742, USA}
\orcid{0000-0002-2742-3447}

\maketitle

\begin{abstract}
The ability to simulate a fermionic system on a quantum computer is expected to revolutionize chemical engineering, materials design, nuclear physics, to name a few. Thus, optimizing the simulation circuits is of significance in harnessing the power of quantum computers. Here, we address this problem in two aspects. In the fault-tolerant regime, we optimize the $\rzgate$ and $\tgate$ gate counts along with the ancilla qubit counts required, assuming the use of a product-formula algorithm for implementation. We obtain a savings ratio of two in the gate counts and a savings ratio of eleven in the number of ancilla qubits required over the state of the art. In the pre-fault tolerant regime, we optimize the two-qubit gate counts, assuming the use of the variational quantum eigensolver (VQE) approach. Specific to the latter, we present a framework that enables bootstrapping the VQE progression towards the convergence of the ground-state energy of the fermionic system. This framework, based on perturbation theory, is capable of improving the energy estimate at each cycle of the VQE progression, by about a factor of three closer to the known ground-state energy compared to the standard VQE approach in the test-bed, classically-accessible system of the water molecule. The improved energy estimate in turn results in a commensurate level of savings of quantum resources, such as the number of qubits and quantum gates, required to be within a pre-specified tolerance from the known ground-state energy. We also explore a suite of generalized transformations of fermion to qubit operators and show that resource-requirement savings of up to more than $20\%$, in small instances, is possible.
\end{abstract}

\section{Introduction}
\label{sec:intro}

Simulating fermionic matter on a quantum computer has recently been receiving much attention. 
Already available in the literature are various chemistry and nuclear physics simulation results \cite{ChemDemon1,ChemDemon2,ChemDemon3,ChemDemon4,ORNL,ORNL2}, performed across multiple quantum computing platforms, including superconducting \cite{ChemDemon1,ChemDemon3,ORNL} and trapped-ion \cite{ChemDemon2,ChemDemon4,ORNL2} based approaches. 
The spotlight on the simulation of fermionic systems on a quantum computer is not accidental. 
Simulations of these systems are useful for furthering fundamental science and practical engineering \cite{IEEE,NAS}, and quantum computers are expected to enable the quantum simulations of fermions undergoing local interactions \cite{Feynman,Lloyd}, a task believed to be classically difficult to scale.

Broadly speaking, simulations of fermionic systems on a quantum computer may be classified into two categories: a variational, quantum-classical hybrid simulation \cite{VQE}, suitable for imperfect, pre-fault tolerant (pre-FT) quantum computers, and a Hamiltonian dynamics simulation based on pure quantum simulation algorithms \cite{PNAS}, typically considered in fault-tolerant (FT) quantum computers. In the context of estimating the ground-state energy of a fermionic system, the former leverages efficient preparation of ansatz states and evaluation of operator expectation values, both enabled by quantum computers. The latter leverages the ability of a quantum computer to efficiently simulate evolution of quantum systems with a local Hamiltonian, which, when combined with quantum phase estimation \cite{Whitfield}, allows us to evaluate the ground-state energy of the system. 

In the pre-FT regime, quantum computational cost is dominated by the use of multi-qubit gates. In the FT regime, where quantum circuits are typically composed of gates in the Clifford+$\tgate$ gateset, quantum computational cost is dominated by the use of $\tgate := \left(\begin{smallmatrix} 1 & 0 \\ 0 & e^{i\pi/4}\end{smallmatrix}\right)$ gates, many of which are used in the FT implementation of $\rzgate(\theta) := \left(\begin{smallmatrix} e^{-i\theta/2} & 0 \\ 0 & e^{i\theta/2}\end{smallmatrix}\right)$ gates (see, e.g., \cite{ar:brs}). 
In this paper, we present approaches that optimize quantum simulations of fermionic systems in either the pre-FT regime, reducing the number of multi-qubit gates, or in the FT regime, reducing the number of $\rzgate$ gates or the number of time steps that $\rzgate$ gates need to be applied, in the simulation circuit.

Our paper is structured as follows. 
In Sec.~\ref{sec:prob} we briefly define the scope of the problem in both the pre-FT and the FT regimes and lay out the preliminaries.
In Sec.~\ref{sec:FT}, we present a highly-optimized circuit construction for the FT quantum simulations, where we improve $\rzgate$ and $\tgate$ counts, as well as the number of ancilla qubits required over the state of the art.
In Sec.~\ref{sec:pFT}, we show two complementary approaches that result in quantum resource savings in the pre-FT regime, i.e., a second-order perturbation based approach and a generalized fermion to qubit operator transformation. 
In the specific sample case of a water molecule, our method provides nearly 25\% improvement over that obtained via the method detailed in \cite{ChemDemon4} in the number of two-qubit gates required to achieve the accuracy to within chemical accuracy.
We discuss our results in Sec.~\ref{sec:disc} and conclude in Sec.~\ref{sec:conc}.

\section{Preliminaries}
\label{sec:prob}

We consider in this paper, as a concrete example, a fermionic system evolving according to a time-independent, local, second-quantized Hamiltonian $H$ in the occupation basis
\begin{equation}
H = \sum_{p,r} h_{pr} a^\dagger_p a_r + \sum_{p,q,r,s} h_{pqrs} a^\dagger_p a^\dagger_q a_r a_s\;,
\label{eq:2ndH}
\end{equation}
where $a^\dagger_p$ and $a_r$ denote the fermionic creation and annihilation operators on the $p$th and $r$th levels, respectively, and $h_{pr}$ and $h_{pqrs}$ denote single- and double- fermion Hamiltonian coefficients, respectively. The fermionic operators follow the canonical anti-commutation relations
\begin{align}
&\{a_j, a_k\} =0\;,{\ \ \ \ }\{a_j^\dagger, a_k^\dagger\}=0\;, {\ \ \ \ }\{a_j,  a_k^\dagger\}=\delta_{jk}\mathbf{1}\;,
\label{anticom}
\end{align}
where $\{A,B\}$ denotes the anti-commutator $AB+BA$, $\delta_{jk}$ is the Kronecker delta, and $\mathbf{1}$ is the identity operator. For their implementations on a quantum computer, the fermionic operators need to be suitably transformed \cite{JW, BK, GM}. A well-known, popular choice is the Jordan-Wigner (JW) transformation \cite{JW}, defined for an $n$-qubit system according to
\begin{equation}
\label{eq:JW}
a_j^\dagger  = 1^{\otimes n - j - 1} \otimes \sigma_+ \otimes  \sigma_z^{\otimes j}\;, \;\; a_j  =1^{\otimes n - j - 1} \otimes \sigma_- \otimes   \sigma_z^{\otimes j}\;,
\end{equation}
where $j\in [0,n-1]$, $\otimes$ denotes the tensor product, $\sigma_{+} := \left(\begin{smallmatrix} 0 & 0 \\ 1 & 0\end{smallmatrix}\right)$, $\sigma_{-} := \left(\begin{smallmatrix} 0 & 1 \\ 0 & 0\end{smallmatrix}\right)$, and $\sigma_z := \left(\begin{smallmatrix} 1 & 0 \\ 0 & -1 \end{smallmatrix}\right)$. 

In the case where we consider a quantum dynamics simulation approach more suitable for the FT regime, we aim to implement the evolution operator $U_{\rm evo} = e^{-iHt}$ on a quantum computer, where $H$ is the system Hamiltonian and $t$ is the duration by which we desire to evolve the system forward in time. 
Roughly, once the initial state $|\psi_{\rm init}\rangle$, sufficiently close to the ground state $|\psi_{\rm ground}\rangle$, is evolved to $|\psi(t)\rangle \approx e^{-iE_{\rm ground}t}|\psi_{\rm init}\rangle$ up to the closeness, quantum Fourier transform may be used to estimate the phase angle $E_{\rm ground}t$, thus extracting the ground-state energy $E_{\rm ground}$. Detailed discussion on the circuit layout that implements the quantum phase estimation algorithm is available in~\cite{NC}. Note, an efficient implementation of the quantum Fourier transform on a FT quantum computer is known~\cite{Nam_QFT}. Therefore, the problem of an efficient energy-spectra calculation on a quantum computer in the FT regime boils down to the problem of efficiently implementing real-time quantum dynamics simulations.

A host of algorithms that (approximately) implement $U_{\rm evo}$ have been proposed, such as the asymptotically optimal quantum signal processing \cite{QSP,QSP2}.
However, it has been shown that for certain Hamiltonians including ones of the form (\ref{eq:2ndH}) a more straightforward technique of the $2k$th order product formula (PF)
can be implemented more efficiently in practice \cite{PNAS,PF}.
Thus in this paper we choose the PF algorithm for the FT regime quantum simulation and present methods to reduce the quantum resources required in its quantum circuit construction, measured in $\rzgate$ gate counts and depths, in addition to the number of $\tgate$ gates and the ancilla qubits required (see Sec.~\ref{sec:FT}).

In the case where we consider a quantum-classical hybrid approach more suitable for the pre-FT regime, 
we consider the widely-adopted technique of the variational quantum eigensolver (VQE) \cite{VQE}. 
Specifically, we aim to implement 
a unitary ansatz evolution operator $U_{\rm ansatz}$ on a quantum computer, 
an example of which is the well-established unitary coupled cluster (UCC) 
ansatz \cite{UCC1,UCC2}. By tuning the variational parameters in $U_{\rm ansatz}$, 
in a typical VQE approach, $\langle \psi_0| U_{\rm ansatz}^\dagger H U_{\rm ansatz} |\psi_0\rangle$ is minimized, 
where $|\psi_0\rangle$ is an initial state that is assumed 
to be close to the target ground state of $H$ and can readily be prepared on a quantum computer. 
The goal of this hybrid approach is then to estimate the ground state energy of the fermionic system with the Hamiltonian of the form (\ref{eq:2ndH}). 
We show in this paper a complementary approach, based on perturbation theory, to the hybrid method described above in order to better and more efficiently estimate the ground state energy (see Sec.~\ref{subsec:perturbation}). 
Our perturbative approach induces a negligible quantum resource overhead, measured in the number of multi-qubit gates. 
We also consider different fermion to qubit transformations other than the aforementioned JW transformation (see Sec.~\ref{subsec:GM}), which help reduce the number of multi-qubit gates without any loss in the algorithmic accuracy.

\section{Fault-Tolerant Regime - Time evolution simulation} 
\label{sec:FT}

As discussed in Sec.~\ref{sec:prob}, in this section, we detail the methods to optimize the time-evolution operator implementation on a quantum computer. In particular, we consider the PF approach. The cost functions we consider are the number of $\rzgate$ gates and the $\rzgate$ gate depth, which are expected to be good proxy measures for the FT-regime quantum resource requirements. 
This is so since the de-facto standard gate set for the FT regime contains $\cnotgate$, single-qubit Clifford, and $\tgate$ gates, where $\tgate$ gates are assumed to be expensive as each implementation of the gate typically requires expending a carefully induced, distilled quantum state that is expensive to produce, and an $\rzgate$ gate consumes multiple $\tgate$ gates for its precise implementation.

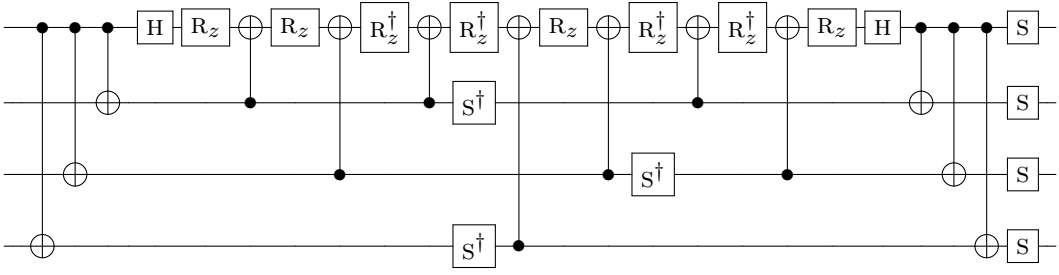
\begin{figure}
\[
\mbox{\Qcircuit @C=0.3em @R=1.0em{
&\qw &\qw &\ctrl{3} &\ctrl{2} &\ctrl{1} &\qw &\gate{\hgate} &\gate{\rzgate} &\targ    &\gate{\rzgate} &\targ    &\gate{\rzgate^\dagger}    &\targ      &\gate{\rzgate^\dagger} &\targ     &\gate{\rzgate}   &\targ      &\gate{\rzgate^\dagger} &\targ    &\gate{\rzgate^\dagger} &\targ        &\gate{\rzgate} &\gate{\hgate}  &\ctrl{1} &\ctrl{2} &\ctrl{3} &\gate{\sgate}     &\qw  &\qw  \\
&\qw &\qw &\qw      &\qw      &\targ    &\qw &\qw           &\qw            &\ctrl{-1}&\qw            &\qw      &\qw                       &\ctrl{-1}  &\gate{\sgate^\dagger}  &\qw       &\qw              &\qw        &\qw                    &\ctrl{-1}&\qw                    &\qw          &\qw            &\qw            &\targ    &\qw      &\qw      &\gate{\sgate}     &\qw  &\qw \\
&\qw &\qw &\qw      &\targ    &\qw      &\qw &\qw           &\qw            &\qw      &\qw            &\ctrl{-2}&\qw                       &\qw        &\qw                    &\qw       &\qw              &\ctrl{-2}  &\gate{\sgate^\dagger}  &\qw      &\qw                    &\ctrl{-2}    &\qw            &\qw            &\qw      &\targ    &\qw      &\gate{\sgate}     &\qw  &\qw \\
&\qw &\qw &\targ    &\qw      &\qw      &\qw &\qw           &\qw            &\qw      &\qw            &\qw      &\qw                       &\qw        &\gate{\sgate^\dagger}  &\ctrl{-3} &\qw              &\qw        &\qw                    &\qw      &\qw                    &\qw          &\qw            &\qw            &\qw      &\qw      &\targ    &\gate{\sgate}     &\qw  &\qw \\}
}
\]
\caption{Standard two-body interaction circuit that implements $\exp[-i(\theta/2) \sigma_+\otimes\sigma_+\otimes\sigma_-\otimes\sigma_- + \mathrm{h.c.}]$. Following the steps detailed in \cite{ChemDemon4} closely, expanding $\sigma_+ \sigma_+ \sigma_- \sigma_- + \mathrm{h.c.}$ (we suppress $\otimes$ hereafter whenever contextually clear) into the particular ordering of $\sigma_x\sigma_x\sigma_x\sigma_x$, $\sigma_x\sigma_x\sigma_y\sigma_y$, $\sigma_x\sigma_y\sigma_y\sigma_x$, $\sigma_x\sigma_y\sigma_x\sigma_y$, $\sigma_y\sigma_y\sigma_x\sigma_x$,
$\sigma_y\sigma_x\sigma_x\sigma_y$,
$\sigma_y\sigma_x\sigma_y\sigma_x$, and
$\sigma_y\sigma_y\sigma_y\sigma_y$, and implementing them one after another with the last qubit as the target qubit, we obtain the circuit shown in this figure after applying the circuit optimization routines detailed in \cite{Optimizer,ChemDemon4}. 
}
\label{fig:standard}
\end{figure}

For completeness, we start by introducing the $2k$th order PF algorithm, simulating the fermionic system described in \sec{prob}. Assuming the JW transformation has been performed on the Hamiltonian $H$ in \eq{2ndH}, the time evolution operator we aim to implement may be written as
\begin{equation}
\exp\biggl(-i\sum_{j=1}^{L}\theta_j\hat{\sigma}^{(j)}\biggr) \approx [S_{2k}(\lambda)]^{r},
\label{eq:PF}
\end{equation}
where $\lambda := 1/r$, $\hat{\sigma}^{(j)} = \bigotimes_i \tilde{\sigma}^{(i,j)} + \text{h.c.}$, where $\tilde{\sigma}^{(i,j)} \in \{\sigma_+, \sigma_-, \sigma_z\}$ and h.c. denotes the Hermitian conjugate operator, and
\begin{align}
S_{1}(\lambda)&:=\prod_{j=1}^{L}\exp(-i\theta_j \hat{\sigma}^{(j)}\lambda), \nonumber \\
S_{2}(\lambda)&:=\prod_{j=1}^{L}\exp(-i\theta_j\hat{\sigma}^{(j)}\lambda/2)\prod_{j=L}^{1}\exp(-i\theta_j\hat{\sigma}^{(j)}\lambda/2), \nonumber \\
S_{2k}(\lambda)&:=[S_{2k-2}(p_{k}\lambda)]^{2}S_{2k-2}((1-4p_{k})\lambda)[S_{2k-2}(p_{k}\lambda)]^{2},
\label{eq:recursive_def}
\end{align}
with $p_{k}:=1/(4-4^{1/(2k-1)})$ for $k>1$ \cite{ar:Suzuki}. 
Inspecting (\ref{eq:PF}) and (\ref{eq:recursive_def}), we observe that the individual exponential terms, hereafter referred to as a Trotter term, are of the form $\exp\left[-i\theta'_j(\bigotimes_i \tilde{\sigma}^{(i,j)} + \text{h.c.} )\right]$, where $\theta'_j$ is a suitably scaled $\theta_j$. A standard circuit that implements a Trotter term is readily available in \cite{Whitfield}. 
Employing the circuit tricks in \cite{ChemDemon4}, in \fig{standard}, we show an optimized quantum circuit that implements an example Trotter term $e^{-i\theta/2(\sigma_{+}\sigma_{+}\sigma_{-}\sigma_{-} + {\rm h.c.})}$.

\begin{figure}
\[
\Qcircuit @C=0.4em @R=1.8em {
	+\; & &\qw       &\ctrl{2} &\ctrl{3} &\gate{\rxgate(\theta)} &\ctrl{3} &\ctrl{2} &\qw        &\qw \\
	+\; & &\targ     &\qw      &\qw      &\ctrl{-1}                      &\qw      &\qw     &\targ      &\qw \\
	-\; & &\ctrl{-1} &\targ     &\qw      &\ctrl{-1}                      &\qw      &\targ   &\ctrl{-1} &\qw \\
	-\; & &\qw       &\qw      &\targ    &\ctrl{-1}                       &\targ    &\qw     &\qw      &\qw
}
\]
\caption{FT-regime optimized circuit for the two-body term $e^{-i\theta/2(\sigma_{+}\sigma_{+}\sigma_{-}\sigma_{-} + {\rm h.c.})}$.
We marked each qubit lines with either $+$ or $-$ to denote which qubits are associated with $\sigma_+$ or $\sigma_-$, respectively, for this particular example.
}
\label{fig:cirq-FT-opt}
\end{figure}
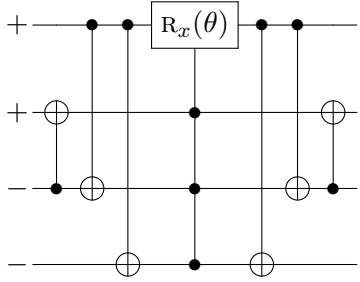

 To more efficiently implement the two-body term in terms of $\rzgate$ counts, we first directly evaluate the operator. It is straightforward to see that the two-body term $e^{-i\theta/2(\sigma_{+}\sigma_{+}\sigma_{-}\sigma_{-} + {\rm h.c.})}$ implements an $\rxgate$-rotation between $\ket{0011}$ state and $\ket{1100}$ state. Note that, for an input sate of $\ket{b_0 b_1 b_2 b_3}$, where the $m$th qubit variable $b_m$ is either 0 or 1, $\ket{b_0\ \ b_1\oplus b_2\ \ b_0\oplus b _2\  \ b_0\oplus b_3}$ maps $\ket{0011}$ to $\ket{0111}$ and $\ket{1100}$ to $\ket{1111}$. A triply-controlled $\rxgate$ rotation on the zeroth qubit would thus implement the desired rotation on the mapped state, and all that needs to be done at this point is to map the rotated state back. We show a quantum circuit that implements the two-body term, constructed according to the aforementioned method, in \fig{cirq-FT-opt}.

We refer to Supplementary Material (SM) Sec.~\ref{app:FTcirc} for the circuit-level implementation details of \fig{cirq-FT-opt}. Briefly, a triply-controlled $\rxgate$ gate in \fig{cirq-FT-opt} may be implemented with Hadamard, $\rzgate$, and triply-controlled $\notgate$ gates.
Specifically, we choose to use relative-phase triply-controlled \notgate\ gates 
(see \cite{RTOF} and SM Sec.~\ref{app:FTcirc} 
for its decomposition into single- and two-qubit gates), as it admits less expensive circuit construction.
Overall, our circuit 
decreases the $\rzgate$ count required per two-body Trotter term 
to two.

Our construction can be contrasted to \cite{Elucidating} wherein (see Fig.~6 of the appendix of \cite{Elucidating}) an optimized implementation of a generic two-body term for the FT regime is discussed. Because the construction in \cite{Elucidating} considers $\sigma_{x/y}$-decomposed form of a two-body term with distinct coefficients -- note our two-body term $e^{-i\theta/2(\sigma_{+}\sigma_{+}\sigma_{-}\sigma_{-} + {\rm h.c.})}$ will have the same coefficients upon decomposition -- the circuit construction there requires eight parallel $\rzgate$ gates with different angles of rotation that are commensurate to the different coefficients. Considering our specific example, these eight angles would be made the same, and the parallel application of the same angle $\rzgate$ gates is known to admit a more efficient implementation using the weightsum trick~\cite{Gidney_2018,ar:LowCost}. As detailed in SM Sec.~\ref{app:FTcirc}, employing this trick reduces the gate requirement from the original $8\, \rzgate$ gates to $32\,\tgate + 4\,\rzgate$ gates. The total number of ancilla qubits required is eleven in this approach. These may be compared with $16\,\tgate + 2\,\rzgate$ gates and one ancilla qubit -- note that a relative-phase triply-controlled $\notgate$ does not require an ancilla, unlike a conventional triply-controlled $\notgate$ gate (see SM Sec.~\ref{app:FTcirc}) -- required for our method that is optimized for $\rzgate$ depth. Both of these methods admit an $\rzgate$ depth of one.

We note in passing that our method described above can be viewed as a generalization of the single-body term implementation discussed elsewhere~\cite{singleref1,singleref2}. Indeed, the generalization may be applied to an arbitrary many-body term, should such a term be of interest to quantum simulations of a more complex kind. See SM Sec.~\ref{app:FTcirc} for detail for the single-body term description, inserted for completeness.

\section{Pre-Fault Tolerant Regime - Variational Quantum Eigensolver} 
\label{sec:pFT}

In the pre-FT regime, we aim to calculate the ground state energy of the system whose Hamiltonian is given in the form (\ref{eq:2ndH}).
This is typically achieved by the VQE approach.
In this approach, by iteratively calling the quantum computer to compute the energy of parametrized ansatz states, one aims to minimize the energy and variationally obtain the ground state of the target system.

We consider in this paper a type of ansatz states $\ket{\Psi_{\rm ansatz}}$
that are transformed from an initial state $\ket{\Psi_0}$ 
using a parametrizable unitary ansatz evolution operator $U_{\rm ansatz}$, 
as $\ket{\Psi_{\rm ansatz}} = U_{\rm ansatz}\ket{\Psi_0}$.
For brevity, we drop the subscript in $U_{\rm ansatz}$ 
in the following text.
Since $U$ is a unitary operator, it can always be
written in the exponential form $e^{Z-Z^\dagger}$
and $Z$ is in turn parametrized.
By varying the parameters in $Z$, the ansatz state energy
$\bra{\Psi_{\rm ansatz}} H \ket{\Psi_{\rm ansatz}}=
\bra{\Psi_0} U^\dagger H U \ket{\Psi_0}$ is
variationally minimized.
With a proper fermion to qubit basis transformation the energy expectation value can be evaluated efficiently on a quantum computer, where the quantum resource cost of this hybrid approach is the implementation cost  of the operator $U$.
We note that the procedures proposed in this section can be implemented 
with any ansatz with a unitary evolution operator $U$.
As a concrete example for discussion, we use
the widely-adopted UCC ansatz with single and double
excitations (UCCSD), a systematic method that is universally 
applicable to any quantum hardware backend. 
Other ansatzes, for instance the $k$-UpCCGSD
ansatz of Lee et al.~\cite{Lee2019} or the
UCCGSD ansatz of Grimsley et al.~\cite{ADAPT-VQE}, 
can also be readily applied.
 
The UCCSD method 
starts with a ground state Hartree-Fock (HF) wavefunction $|\Psi_0\rangle$ 
that can be easily calculated on a classical computer 
and implemented on a quantum computer. 
It then evolves $|\Psi_0\rangle$ with a unitary operator $U= e^{Z_{\rm UCCSD}-Z_{\rm UCCSD}^\dagger}$, 
where $Z_{\rm UCCSD} = \sum_{l=1}^2 Z_l$ is the
so-called cluster operator. 
$Z_{1,2}$ are the single and double excitation operators 
and are written in second quantization as
\begin{align}
\begin{split}
 Z_1 =& \sum_{\substack{p\in \mathrm{virt} \\ r\in \mathrm{occ}}} t_{pr} a^\dagger_p a_r, \\
 Z_2 =& \sum_{\substack{p,q\in \mathrm{virt} \\ r,s\in \mathrm{occ}}} t_{pqrs} a^\dagger_p a^\dagger_q a_r a_s,
\end{split}
\label{eqn:ansatz}
\end{align}
where $t_{pr}$ and $t_{pqrs}$ are variational parameters and ``virt'' and ``occ'' denote virtual and occupied levels respectively.
Notice that for the UCCSD ansatz, to minimize
the size of the circuit that implements $U$, 
not all terms in the cluster operator are needed to achieve a certain precision. 
In other words, only a subset of all the parameters
$t_{pq}$ and $t_{pqrs}$ need to be included in the VQE approach,
while the rest can be set to zero.
In fact, it is very much the name of the game in the
pre-FT regime to find a minimal set of variational 
parameters so that the final ground state energy
satisfies certain error criteria.

In this section,
we propose two general procedures to reduce
circuit complexity of the approach detailed above,
represented by the total number of multi-qubit gates.
To be concrete with our example, we use the first-order PF algorithm to implement the UCCSD ansatz, although extensions to other ansatzes or higher order PF algorithms are straightforward.
In Sec.~\ref{subsec:perturbation}, we detail a hybrid framework that is based on perturbation theory, wherein we apply a perturbation correction to each instance of the VQE circuit. As will be discussed in Sec.~\ref{subsec:perturbation} in detail, our perturbation approach naturally provides a means to determine the variational parameters to further include in the ansatz for a given VQE circuit. This enables us to systematically build, in an iterative fashion, a larger and larger VQE circuit, effectively bootstrapping the VQE progression towards the ground-state energy estimate of the simulated system. 
In Sec.~\ref{subsec:GM}, we explore a suite of generalized fermion to qubit operator transformations to reduce the cost of quantum computation without sacrificing algorithmic accuracy.

\subsection{Perturbation Assisted Quantum Simulation} 
\label{subsec:perturbation}

In this section, we describe a general framework that leverages the
power of perturbation theory to optimize VQE-based 
quantum simulations by predicting the
subset of variational parameters
to include in the ansatz
and correcting the VQE result via post processing.
The framework optimizes both the total number of VQE executions
as well as the size of the ansatz state preparation circuits used to reach convergence in the ground-state energy estimate.
In Sec.~\ref{sssec:prtb_framework} we outline the framework. 
The derivation of a simple perturbation scheme that can be
straightforwardly implemented in the framework is
shown in Sec.~\ref{sssec:prtb_hmp2}, which we hereafter refer to as
a hybrid second order M{\o}ller-Plesset perturbation 
(HMP2) method. In Sec.~\ref{sssec:prtb_data}
we present a classically simulated comparison between the
HMP2 scheme 
and a more conventional VQE approach~\cite{ChemDemon4}, 
by computing the ground state energy of a water molecule
using the UCCSD ansatz.

\subsubsection{Perturbative predictor and corrector}
\label{sssec:prtb_framework}

\begin{figure}
\centering
    \includegraphics[width=0.95\columnwidth]{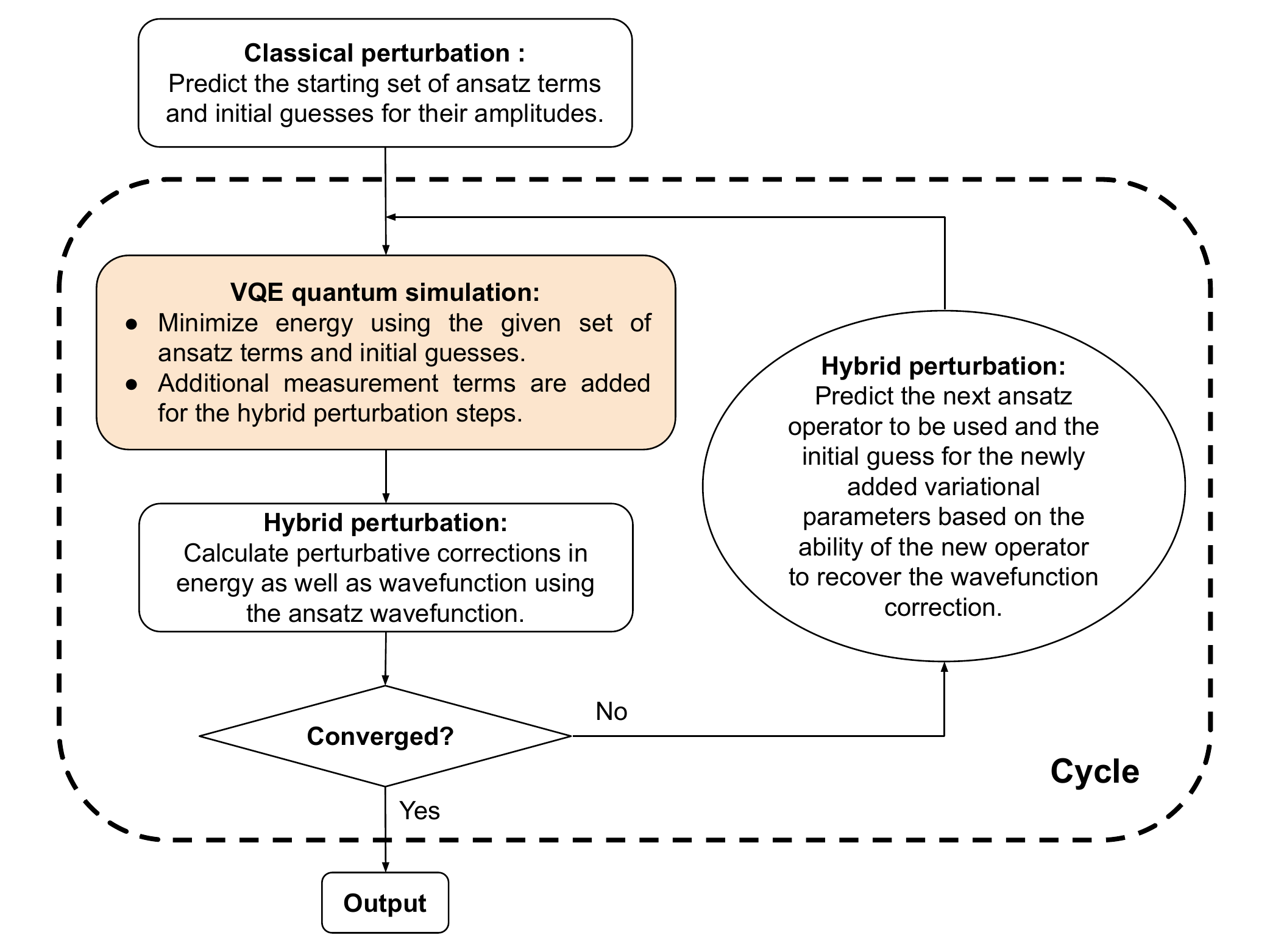}
    \caption{Flow diagram for the proposed framework incorporating
        perturbation methods to VQE simulations. See main text for detailed
        description.}
    \label{fig:pflow}
\end{figure}

We detail, in this section, a general, systematic framework of using perturbation methods to improve quantum resources required for the VQE approach. 
Specifically, we aim to rapidly converge to the ground state energy with small quantum circuits.

Any general (not specific to any system) unitary ansatz
operator can be written as
\begin{equation}
    U = e^{Z - Z^\dagger} \;, \; Z = \sum_{\{\alpha\}} 
      f_{\alpha}(\{t_\beta\}) D_{\alpha},
      \label{eq:generalU}
\end{equation}
where the summation is over a set of orbital sequences $\{\alpha\}$, 
$f_{\alpha}$ are real functions of the set of variational
parameters $\{t_\beta\}$, and
$D_{\alpha}$ is a generic orbital substitution operator
that substitutes the orbitals in a wavefunction according to 
the orbital sequence $\alpha$. 
For instance, a single orbital substitution operator 
$D_{p^\dagger q} = a_p^\dagger a_q$ substitutes orbital $q$ with $p$.
We can then define the orbital substituted wavefunction
$\ket{\Psi_{D_\alpha}} = D_{\alpha}\ket{\Psi_0}$.
For a specific general ansatz approach there is a full set of variational parameters $T_{\rm full} = \{t_\beta\}$
that can be used to parametrize the ansatz evolution operator.
In the case of UCCSD, by comparing (\ref{eqn:ansatz})
with (\ref{eq:generalU}), we note that
$T_{\rm full} = \{t_{pr}, t_{pqrs} ~ | ~ p,q\in {\rm virt}; ~ r,s\in {\rm occ}\}$,
$\{\alpha\} = \{p^\dagger r, p^\dagger q^\dagger rs ~ | ~ p,q\in {\rm virt}; ~ r,s\in {\rm occ}\}$,
and $f_{\alpha} = t_{\alpha}$.

Note further that we can generate a set of ansatz evolution operators $\{U_{T_i}\}$, 
each element of which is the unitary operator that induces an ansatz state
with a subset $T_i$ of the full variational parameters $T_{\rm full}$ ($T_i \subset T_{\rm full}$).
The rest of the parameters in
the complement set $T_{\rm full}\setminus T_i$
are fixed to an ordered set of default values $T_{\rm full}^{(0)}$.
Without loss of generality, we assume the default values
are absorbed into $f_{\alpha}$ in (\ref{eq:generalU})
and the elements in the complement set $T_{\rm full}\setminus T_i$
are simply set to zero.
For every $U_{T_i}$ in $\{U_{T_i}\}$, the ansatz state energy
can be variationally minimized as 
$E_{T_i} = \mathop{\boldsymbol\min}_{{\rm elements}(T_i)\in\mathbb{R}} 
\bra{\Psi_0} U_{T_i}^\dagger H U_{T_i} \ket{\Psi_0}$,
where ${\rm elements}(T_i)$ denotes the elements of the
set $T_i$.
For an application of a particular general ansatz
through the VQE approach in the pre-FT regime,
one aims to find a $U_{T_i}$ that has the
least circuit complexity while satisfying 
$|E_{T_i} - E_{T_{\rm full}}| < \varepsilon$
with $\varepsilon$ a pre-determined error criterion.

A concrete strategy to reach convergence with minimal
circuit complexity is to start with an operator with a small 
set of variational parameters that is known to be necessary
(but not necessarily sufficient) to achieve the error threshold and iteratively grow it.
Going from the $m$th iteration to the $(m+1)$th one, 
the set of variational parameters in the operators grows and
satisfies $T_m\subset T_{m+1}$.
A critical part in ensuring a rapid convergence is the selection of 
the additional variational parameter(s) to include between
iterations.
In the case of UCCSD, they directly correspond to additional
individual excitation terms in (\ref{eqn:ansatz}) used to prepare the ansatz state.
Previous works have, e.g., used the full configuration interaction (FCI) results \cite{ChemDemon4} in the case the system is sufficiently small to be simulatable on a classical computer, and have demonstrated the significance of the excitation term selection for the resource requirement. Here, we propose to iteratively select the next variational parameters to include in the ansatz state preparation based on the size of the
perturbatively predicted wavefunction correction amplitude 
for a given ansatz state whose variational parameters are already optimized via the conventional means of VQE. Our strategy also evaluates a perturbative energy correction in addition to the conventional VQE approach, directly contributing to the fast convergence, while incurring no overhead in the pre-FT regime resource requirement.

Figure~\ref{fig:pflow} shows the flow diagram of the
proposed framework. As in the conventional VQE approach 
for fermionic simulations, we start with the ground state
Hartree-Fock wavefunction 
of the single particle Hamiltonian as $\ket{\Psi_0}$. 
The first step is to determine the initial evolution
operator $U_{T_1}$ to use for the first iteration of VQE.
As an example, using the two-particle Hamiltonian as perturbation, energy 
and wavefunction corrections can be calculated using 
classical algorithms with relative ease. 
From the perturbed wavefunction,
we can extract the amplitudes of individual excited states in the single-particle Hamiltonian basis. 
These amplitudes will serve as the initial guesses of the variational parameters for the first round
of VQE simulation, which could significantly reduce
the number of evaluations of the quantum circuit
comparing to all-zeros or random initial 
guesses~\cite{MP2-ordering}. 
If we demand the energy convergence criteria of $\delta_E$
-- The entire simulation is considered to be
converged when the magnitude of energy change
associated with an addition of any extra ansatz term 
is smaller than $\delta_E$ --
we include in the initial ansatz set the ansatz terms whose contribution to the correlation energy is greater than $f(\delta_E)$, where a standard choice of function $f(\delta_E)$ may simply be $\delta_E$.

With the initial set of ansatz terms and their initial variational parameter values,
    we run the first round of VQE simulation to minimize the energy 
    $E_{T_1}=\bra{\Psi_0} U_{T_1}^\dagger H U_{T_1} \ket{\Psi_0}$.
Once the energy is minimized and the ansatz state converges, we proceed to compute the expectation values of a set of operators that were chosen in advance to inform us about the perturbation around the converged ansatz state.
We refer to this method as a hybrid perturbation since
unlike the conventional perturbation where we know the unperturbed 
wavefunction in advance, we start the perturbation from the converged VQE ansatz state.

Based on the additional measurements performed on a quantum computer,
    we may now use the hybrid perturbation method to calculate
    corrections to the correlation energy and 
    the wavefunction. 
The resulting total energy of this cycle
    is the summation of the energy correction and the
    VQE energy.
Before starting the next cycle, we first identify
a pool of operators that can be the immediate successor
of the current operator. Then we evaluate how much each operator in the pool can account for towards the perturbative wavefunction
correction. The operator that can recover the wavefunction correction the most gets the nod for the next cycle and the way it recovers the correction informs the initial guesses of the
newly added variational parameters in the next cycle. 
A more detailed illustration of such a procedure
is given in the next subsection.
The initial variational parameter values of the ansatz terms 
    that continue to be a part of the next VQE simulation may simply be imported from the cycle before.

The cycle iterates the outlined procedure of running
the VQE simulation and the hybrid perturbation calculation, 
until convergence of the total energy is achieved. We next detail the steps
of a simple implementation of our hybrid perturbation framework that is inspired by the classical MP2 method.

\subsubsection{HMP2 method}
\label{sssec:prtb_hmp2}

We derive in this section our hybrid perturbation method, inspired by the M{\o}ller-Plesset
    perturbation theory, applied to the VQE simulation.
Our hybrid MP2-based perturbation method, HMP2, aims
to help each VQE cycle in terms of both
improving the resulting energy via a perturbative energy correction
and optimizing the ansatz operator to use in the
next cycle through a perturbative wavefunction correction.  

In the $m$th VQE cycle, we can write
the energy of the ansatz state $E_{T_m}$
in the context of first-order perturbation theory, i.e.,
\begin{align}
    \begin{split}
        E_{T_m} &= \bra{\Psi_0}  U_{T_m}^\dagger H U_{T_m} \ket{\Psi_0 } \\
    		    &= \bra{\Psi_0}  F + (U_{T_m}^\dagger H U_{T_m} - F) \ket{\Psi_0 } \\
    		    &= \bra{\Psi_0} F \ket{\Psi_0 }  + \bra{\Psi_0} U_{T_m}^\dagger H U_{T_m} - F \ket{\Psi_0 } \\
    		    &= E_0 + (E_{T_m}- E_0)\\
    		    &= E^{(0)} + E^{(1)}_{T_m},
    \end{split}
    \label{eq:EU}
\end{align}
where $F$ is the Fock operator, $E_0$ is the sum of orbital energies,
$E^{(0)} = E_0$ is the zeroth-order energy,
and $E^{(1)}_{T_m} = E_{T_m} - E_0 $ 
is the first-order correction energy.
Based on perturbation theory, a second-order correction to the energy can now be written as
\begin{equation}
    \label{eqn:2nd-perturb}
    E^{(2)}_{T_m} =   \sum_{\{\alpha^\prime\}}\frac{\left|\bra{\Psi_{D_{\alpha^\prime}}} 
                V_{T_m} \ket{\Psi_0\vph}\right|^2}{\Delta E_{D_{\alpha^\prime}}},
\end{equation}
where $V_{T_m} = U_{T_m}^\dagger H U_{T_m} - F$.
The set of orbital sequences $\{\alpha^\prime\}$ does not necessarily need to
be the same as the set $\{\alpha\}$ used in (\ref{eq:generalU}) and should
be chosen according to accuracy requirement and resource constraints.
See Sec.~\ref{sec:disc} for more detailed discussion.
In our case study using the UCCSD ansatz, 
we choose $\{\alpha^\prime\} = \{\alpha\}$.
The energy $\Delta E_{D_{\alpha^\prime}}$ is defined as the
    orbital energy difference of the orbital substitutions. 
For instance, $\Delta E_{D_{p^\dagger q}} = E_q - E_p$, 
    where $E_{p}$ and $E_{q}$ 
    are the orbital energies of the $p$th and the $q$th orbitals, respectively.  
Inserting  $V_{T_m} = U_{T_m}^\dagger H U_{T_m} - F$ in (\ref{eqn:2nd-perturb}), 
    the numerator becomes
\begin{align} \label{eqn:perturb-exact}
\begin{split}
\left|\bra{\Psi_{D_{\alpha^\prime}}} V_{T_m} \ket{\Psi_0\vph}\right|^2 
&=\left|\bra{\Psi_{D_{\alpha^\prime}}} U_{T_m}^\dagger H U_{T_m} - F \ket{\Psi_0\vph}\right|^2\\
&=\left|\bra{\Psi_{D_{\alpha^\prime}}} U_{T_m}^\dagger H U_{T_m} \ket{\Psi_0\vph} - \bra{\Psi_{D_{\alpha^\prime}}} F \ket{\Psi_0\vph}\right|^2 \\
&=\left|\bra{\Psi_{D_{\alpha^\prime}}} U_{T_m}^\dagger H U_{T_m} \ket{\Psi_0\vph}\right|^2,
\end{split}
\end{align}
where we used 
$\bra{\Psi_{D_{\alpha^\prime}}} F \ket{\Psi_0\vph}=0$.

In order to apply the VQE results, especially
the ansatz wavefunction $\ket{\Psi_{\rm ansatz}^{T_m}}$ directly to 
the perturbation calculation, we proceed as follows. 
For brevity, we introduce $\tilde{Z}_{T_m} = Z_{T_m} - Z_{T_m}^\dagger$ with 
$U_{T_m} = e^{\tilde{Z}_{T_m}}$ and $\tilde{D}_{\alpha^\prime} = D_{\alpha^\prime}-D_{\alpha^\prime}^\dagger$.
Inserting these into (\ref{eqn:perturb-exact}), we obtain
\begin{align} 
\label{eqn:perturb-rewrite}
\begin{split}
|\bra{\Psi_{D_{\alpha^\prime}}} U_{T_m}^\dagger H U_{T_m} \ket{\Psi_0}|^2
& = |\bra{\Psi_0}\tilde{D}_{\alpha^\prime}^\dagger  U_{T_m}^\dagger H U_{T_m} \ket{\Psi_0}|^2 \\
& = |\bra{\Psi_{\rm ansatz}^{T_m}} U_{T_m} \tilde{D}_{\alpha^\prime}^\dagger  U_{T_m}^\dagger H \ket{\Psi_{\rm ansatz}^{T_m}}|^2 
\;.
\end{split}
\end{align}
Next we Taylor expand the $e^{\pm \tilde{Z}_{T_m}}$ in
$e^{\tilde{Z}_{T_m}} \tilde{D}_{\alpha^\prime}^\dagger  e^{-\tilde{Z}_{T_m}} H$ up to
first order in $\tilde{Z}_{T_m}$. This is consistent with our
choice of the first-order PF algorithm for the implementation
of the ansatz. We obtain
\begin{align} 
\label{eqn:perturb-approx}
\begin{split}
&|\bra{\Psi_{\rm ansatz}^{T_m}} e^{\tilde{Z}_{T_m}} \tilde{D}_{\alpha^\prime}^\dagger  e^{-\tilde{Z}_{T_m}} H \ket{\Psi_{\rm ansatz}^{T_m}}|^2 \\
= &|\bra{\Psi_{\rm ansatz}^{T_m}}  (1+\tilde{Z}_{T_m}+\cdots) \tilde{D}_{\alpha^\prime}^\dagger  (1-\tilde{Z}_{T_m}+\cdots) H  \ket{\Psi_{\rm ansatz}^{T_m}}|^2\\
\approx &|\bra{\Psi_{\rm ansatz}^{T_m}} \tilde{D}_{\alpha^\prime}^\dagger H - \tilde{D}_{\alpha^\prime}^\dagger \tilde{Z}_{T_m}H + \tilde{Z}_{T_m}\tilde{D}_{\alpha^\prime}^\dagger H \ket{\Psi_{\rm ansatz}^{T_m}}|^2.
\end{split}
\end{align}
With (\ref{eqn:2nd-perturb}) and (\ref{eqn:perturb-approx}) we can obtain
the energy correction to the VQE result.

We note that $\tilde{D}_{\alpha^\prime}^\dagger H - \tilde{D}_{\alpha^\prime}^\dagger \tilde{Z}_{T_m}H + \tilde{Z}_{T_m}\tilde{D}_{\alpha^\prime}^\dagger H$ is a sum of products of Pauli operators $\sum_j \epsilon_j \hat{\sigma}^{(j)} $ in the qubit basis, after applying a suitable fermion to qubit basis transformation. Since $\hat{\sigma}^{(j)}$ have eigenvalues $+1$ or $-1$,
denoting $\ket{\Psi_j^{(p)}}$ and $\ket{\Psi_j^{(m)}}$ as the eigenvectors with the respective eigenvalues, we may then write 
\begin{align} 
\label{eqn:final}
\begin{split}
&\left | \left \langle \Psi_{\rm ansatz}^{T_m} \left | \sum_j \epsilon_j \hat{\sigma^{(j)}} \right | \Psi_{\rm ansatz}^{T_m} \right \rangle  \right|^2\\ =&
\left| \left \langle \Psi_{\rm ansatz}^{T_m} \left|  \sum_j \epsilon_j \left(\sum_p \ket{\Psi_j^{(p)}}\bra{\Psi_j^{(p)}} - \sum_m \ket{\Psi_j^{(m)}}\bra{\Psi_j^{(m)}}\right) \right | \Psi_{\rm ansatz}^{T_m} \right \rangle  \right|^2 \\
=&\left | \sum_j \epsilon_j \left(\sum_p \left | \left \langle \Psi_{\rm ansatz}^{T_m} \left. \right| \Psi_j^{(p)} \right \rangle \right|^2 - \sum_m \left| \left \langle \Psi_{\rm ansatz}^{T_m} \left. \right|  \Psi_j^{(m)} \right \rangle \right|^2 \right)   \right|^2 \;.
\end{split}
\end{align}
Note that (\ref{eqn:final}) requires only a simple projection of $\ket{\Psi_{\rm ansatz}}$ onto $\ket{\Psi_j^{(p)}}$ or $\ket{\Psi_j^{(m)}}$. Thus the second order correction energy may be obtained without any quantum resource overhead in the circuit size.

To optimize the choice of the ansatz evolution operator to use in the next cycle,
we turn to the wavefunction correction. The ansatz state wavefunction
can be considered as the zeroth-order wavefunction $\ket{\Psi^{(0)}}$ and a
first order correction is given by
\begin{equation}
  \ket{\Psi^{(1)}_{T_m}} = \sum_{\{\alpha\}}
    \frac{\bra{\Psi_{D_{\alpha}}} 
   V_{T_m} \ket{\Psi_0\vph}}{\Delta E_{D_{\alpha}}}
    \ket{\Psi_{D_{\alpha}}}.
\label{eq:HMP2WvfCorr}
\end{equation}
We then use the following 
procedure to determine which ansatz evolution operator to use next in the $(m+1)$th round. 
In the $m$th round we have variationally determined
a set of values for the parameters in the set $T_m$, denoted as $S_m$.
We then look for a set $R_m = \{U_{T_i}\}$ of all the evolution operators
$U_{T_i}$ that satisfies $T_m \subset T_i$ and $n(T_i \setminus T_m) = 
\min_{j\ne m}[n(T_j \setminus T_m)]$, where $n(A)$ denotes the number of
elements in set $A$. $R_m$ is then the pool of operators from which
we are going to choose the next ansatz evolution operator.
For each $U_{T_i}$ in $R_m$, we can then compute the overlap
$F_{m}^{i} = 1/w_{m}^{i}\sum_{t_\beta\in T_i \setminus T_m}\left|\left\langle \Psi_{t_\beta}^{m,i} | \Psi^{(1)}_{T_m}\right\rangle\right|$, where
\begin{equation}
    \ket{\Psi_{t_\beta}^{m,i}} = \left(
    \left.\frac{\partial U_{T_i}}{\partial t_\beta}\right|_{S_m}\right)\ket{\Psi_0} \;.
\label{eq:psimi}
\end{equation}
Here we use $w_{m}^{i}$ to capture the additional cost
associated with implementing $U_{T_i}$ instead of $U_{T_m}$.
For example, it can be the number of additional 
two-qubit gates needed to compute $\bra{\Psi_0} U^\dagger H U \ket{\Psi_0}$
for $U = U_{T_i}$ instead of
$U = U_{T_m}$. For our concrete example, we simply take
$w_{m}^{i}$ to be the number of additional fermionic
operators in $T_i$ comparing to $T_m$.
Using the identity $\left\langle \Psi_{D_\alpha} | \Psi_{D_\alpha^\prime}\right\rangle
=\delta_{\alpha\alpha^\prime}$, it is straightforward to evaluate all of the $F_m^i$ using
(\ref{eqn:2nd-perturb}), (\ref{eqn:perturb-approx}), and (\ref{eq:psimi}).
Then we simply pick the $U_{T_i}$ out of $R_m$ with the largest $F_m^i$
to be the ansatz evolution operator to use in the next cycle.
We can also use the wavefunction correction to guess
the initial values for an additional variational parameter
$t_\beta \in T_{m+1} \setminus T_m$ as $\pm \left\langle \Psi_{t_\beta}^{m,m+1} | \Psi^{(1)}_{T_m}\right\rangle$.

In the case of UCCSD, each $U_{T_i}$ in $R_m$ only has one more $D_\alpha$
term in $Z_{T_i}$ than $Z_{T_m}$ of $U_{T_m}$, which
corresponds to an incremental change in the terms
included in the ansatz operator $Z$. Using $e^{\tilde{Z}_{T_i}} \approx 1 + \tilde{Z}_{T_i}$,
$F_m^i$ is approximately the perturbative amplitude
$|\bra{\Psi_{D_{\alpha}}}V_{T_m} \ket{\Psi_0}/\Delta E_{D_{\alpha}}|$
of the corresponding additional $D_{\alpha}$.

All the procedures introduced above for the $m$th
round of a VQE cycle can also be used before the first cycle
by simply taking $m = 0$. In this case, all the
calculations are classical and the energy
correction is simply the classical MP2 energy correction.
The wavefunction correction however can inform the first
operator to be tried and suggest the initial values 
for its variational parameters.

We note in passing that, in principle, the number of individual Pauli-string operators
in (\ref{eqn:final})
whose expectation values are to be evaluated using a quantum computer scales as $O(n^4)$ per UCCSD ansatz where $n$ is the number of qubits. This is so since $H$ consists of $O(n^4)$ terms, while,
for a particular ansatz, $Z$ is fixed and provides a constant prefactor.
Thus the predictive feature using (\ref{eq:HMP2WvfCorr}) requires
$O(n^8)$ Pauli-string measurements because there are $O(n^4)$
elements in $\{\alpha\}$. The perturbative correction using (\ref{eqn:2nd-perturb})
can reuse all the measurement results from the predictive feature 
because we choose $\{\alpha'\} = \{\alpha\}$.
While this may appear challenging, in practice, a series of techniques can be applied to significantly reduce the number of evaluations. 
First, we can take advantage of the fact that only the qubits that are coupled via a chosen ansatz are entangled. 
By carefully choosing the qubits to use to create the ansatz state, we can treat the qubits that have not been operated on classically. 
A straightforward extension to small, disjoint sets of qubits with set sizes that admit inexpensive classical postprocessing may be considered as well, although we do not consider such an aggressive optimization in this paper.
In addition, if we have two qubits that represent the same spatial orbital with two opposite spins, in the case where they are not distinguishable in their energy due to a particular choice of the ansatz (see Sec.~\ref{subsec:GM} and \cite{ChemDemon4} for more details), the two qubits would encode a redundant piece of information, and this allows us to encode the information using just one qubit.
We call this optimization technique qubit space reduction (QSR) (see SM Sec.~\ref{app:reduce-qubit-space} for the implementation details).
Secondly, it is possible to measure multiple Pauli strings simultaneously to reduce the total number of measurements provided that these Pauli strings commute with each other. 
For instance, two methods,
the general commuting (GC) partition and the
qubit-wise commuting (QWC) partition, have been proposed
to enable such simultaneous measurements~\cite{simu-measure}.
Generally speaking, GC can reduce the number of measurements significantly but could incur additional cost in terms of additional two-qubit gates while
QWC can reduce the number of measurements, though to a lesser degree than GC, without increasing the number of two-qubit gates.
If using GC, it is beneficial to use QSR beforehand to
reduce the size of the qubit space needed to be partitioned,
which in turn reduces the number of extra two-qubit gates incurred from using the GC partition.

\subsubsection{Comparison to prior state-of-the-art}
\label{sssec:prtb_data}

\begin{table}[]
\centering
\begin{tabular}{ccccc}
\multicolumn{2}{c}
{Ansatz}& $E_\mathrm{UCCSD}$ 
        & $E_{\mathrm{P_D}}^{\text{corr}}$ 
        & $E_{\mathrm{UCCSD+HMP2}}$  \\ \hline \hline
\multirow{9}{*}{\begin{tabular}[c]{@{}c@{}}Ansatz \\ terms in\\ HMP2 \\ order\end{tabular}}  
& HF+1  &-74.9749   &-0.0249    &-74.9998  \\
& HF+2  &-74.9781   &-0.0220    &-75.0001  \\
& HF+4  &-74.9854   &-0.0170    &-75.0024  \\
& HF+5  &-74.9881   &-0.0155    &-75.0036  \\
& HF+6  &-74.9909   &-0.0139    &-75.0048  \\
& HF+8  &-74.9966   &-0.0091    &-75.0057  \\
& HF+9  &-74.9996   &-0.0068    &-75.0063  \\
& HF+11 &-75.0038   &-0.0039    &-75.0077  \\
& HF+12 &-75.0047   &-0.0034    &-75.0081  \\
& HF+14 &-75.0074   &-0.0023    &-75.0098  \\
& HF+16 &-75.0091   &-0.0015    &-75.0106 \\
& HF+17 &-75.0100   &-0.0009    &-75.0109 \\
& HF+19 &-75.0102   &-0.0007    &-75.0109 \\
& HF+21 &-75.0107   &-0.0004    &-75.0111\\
& HF+23 &-75.0111   &-0.0002    &-75.0113\\
& HF+24 &-75.0113   &-0.0001    &-75.0114\\
& HF+26 &-75.0113   &-0.0001    &-75.0114 \\
& HF+28 &-75.0113   &-0.0001    &-75.0114 
\\\hline
$E_{\text{HF}}$ &  &  &  &  -74.9624 \\
$E_{\text{MP2}}$ &  &  &  &  -74.9977 \\
$E_{\text{CCSD}}$ &  &  &  &  -75.0114 \\
$E_{\text{FCI}}$ &  &  &  &  -75.0116
\end{tabular}
\caption{Ground state energy calculations for a water
molecule using STO-3G basis. 
HF+$N$ represents a VQE cycle with the UCCSD ansatz with $N$ 
$D_{\alpha}$ terms, as described in 
Sec.~\ref{sssec:prtb_framework}. 
$E_{\mathrm{P_D}}^{\text{corr}}$ is
the HMP2 correction based on (\ref{eqn:perturb-approx}).
$E_{\mathrm{UCCSD+HMP2}} = E_\mathrm{UCCSD} + E_{\mathrm{P_D}}^{\text{corr}}$
is the total energy obtained
in one VQE cycle.
For the classical optimization step in VQE, we used L-BFGS-B optimizer~\cite{zhu1997algorithm}, and we assumed perfect measurement with an infinite number of circuit repetitions to suppress the effect of finite measurement.
The classically computed energies $E_{\text{HF}}$, $E_{\text{MP2}}$, $E_{\text{CCSD}}$, and $E_{\text{FCI}}$
for the water molecule with the same geometry and basis set
are also listed for comparison.
All energies are in units of Hartree.}
\label{tab:hmp2_order}
\end{table}

To demonstrate our framework of perturbation assisted
quantum simulation using the HMP2 method, 
we performed classically emulated VQE calculations
with the UCCSD ansatz
of the ground state energy of a water molecule
at its equilibrium geometry.
Using the STO-3G basis, the 
calculation contains 14 qubits in total.
Table~\ref{tab:hmp2_order} shows the incremental changes
of the UCCSD correlation energy as well as the HMP2
correction as more ansatz terms are included according
to our framework.
Note that our choice of $\{\alpha^\prime\}$ in 
(\ref{eqn:2nd-perturb}) for this example guarantees that
the final energy reaches the UCCSD energy when
all the ansatz terms are included in (\ref{eqn:ansatz}).
To compare, we also listed the classically calculated
HF, MP2, CCSD, and FCI energies since it is
difficult to obtain the UCCSD energy via classical algorithms.
With the HMP2 correction, the total energy
$E_{\text{UCCSD+HMP2}}$ descends quickly towards FCI energy.

\begin{figure}
\centering
\includegraphics{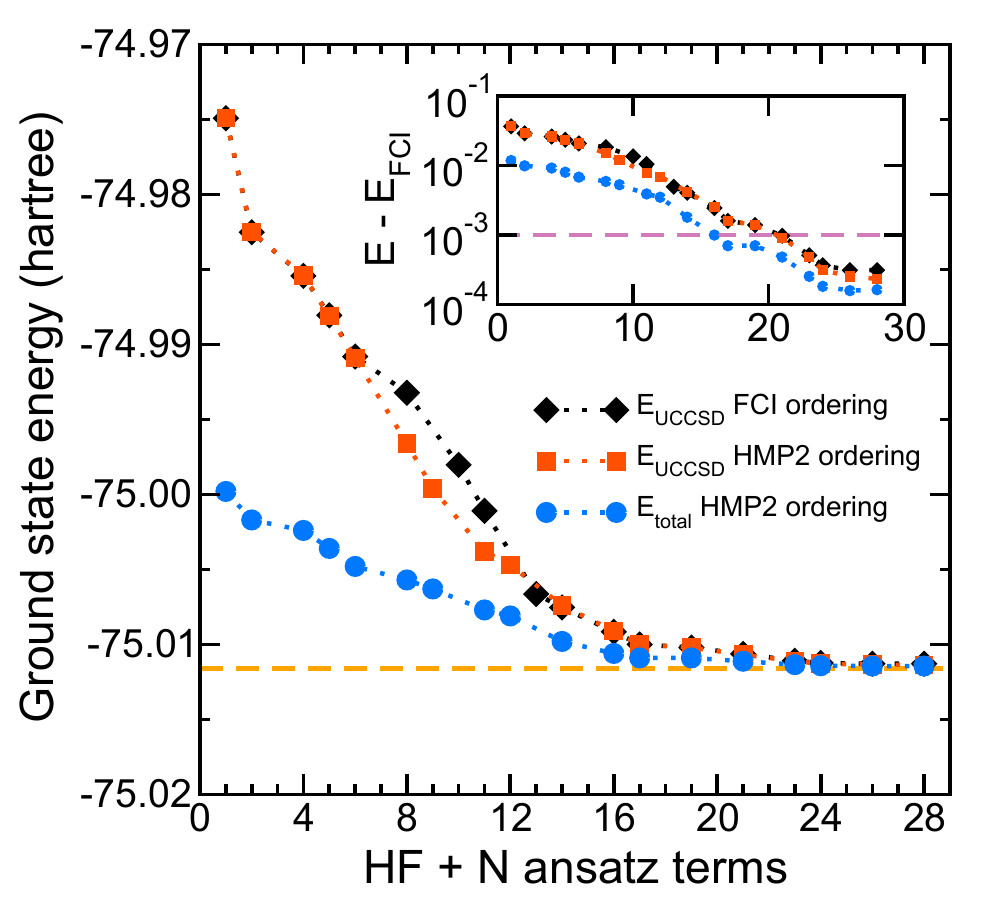}
\caption{Comparison of the ground state energies of a water molecule
at its equilibrium geometry using the STO-3G basis set, calculated
by various methods as a function of $N$, 
the number of $D_{\alpha}$ terms included
in the HF$+N$ ansatz. 
The orange dashed line is the FCI energy
calculated using the PSI4 package~\cite{psi4}, which serves as
the benchmark. The black diamonds connected by the dotted lines
are the UCCSD energies 
$E_\mathrm{UCCSD}$, calculated using different numbers
of ansatz terms ordered by the contribution of corresponding
determinants to the FCI energy. The red squares and blue circles
connected by the dotted lines are the ground-state energies computed according to the proposed
framework with the HMP2 ordering, with the circles containing
the additional HMP2 energy correction at each VQE cycle. 
The inset shows in semi-log the differences between the 
energies obtained by the aforementioned methods and the FCI energy
as a function of $N$. The purple dashed
line shows the chemical accuracy given by $10^{-3}$ hartree.
As stated in the caption of Table~\ref{tab:hmp2_order} as well, for the classical optimization step in VQE, we used L-BFGS-B optimizer~\cite{zhu1997algorithm}, and we assumed perfect measurement with an infinite number of circuit repetitions to suppress the effect of finite measurement.
}
\label{fig:convg}
\end{figure}

%Begin new
\begin{figure}
\centering
\includegraphics[width=0.95\columnwidth]{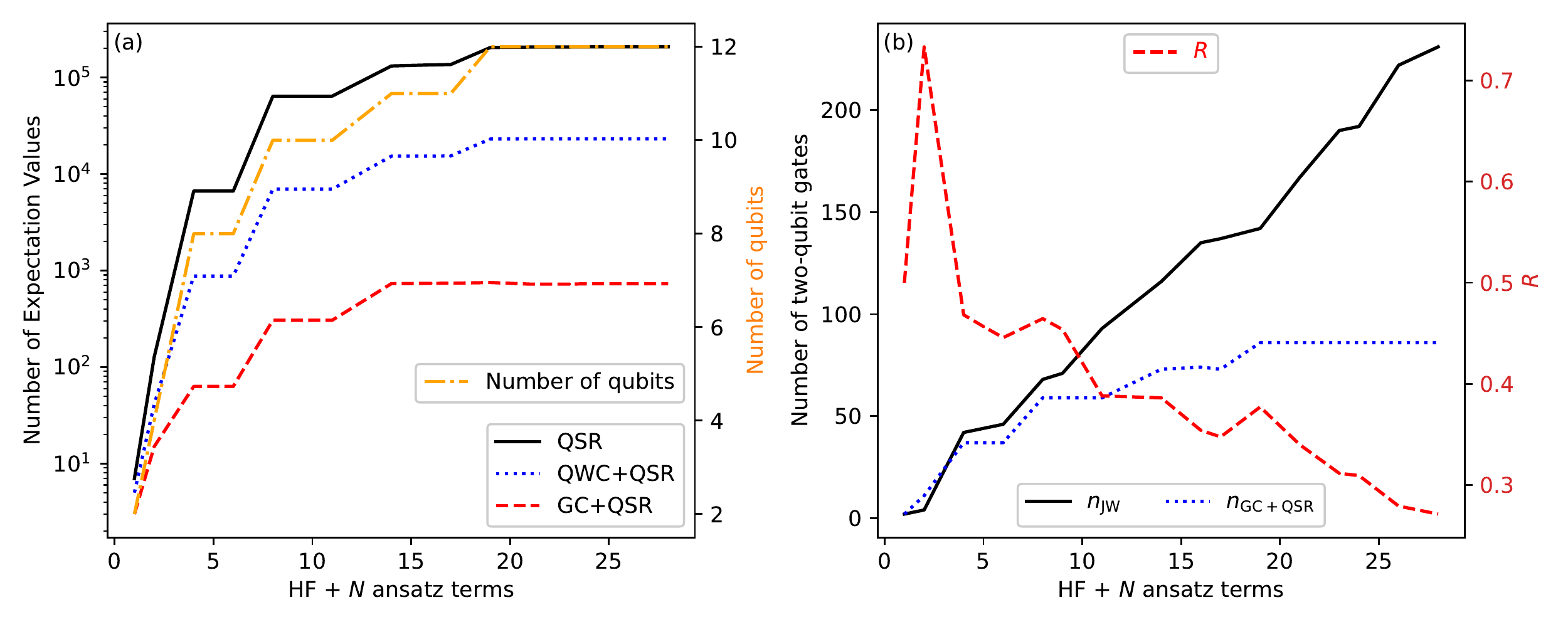}
\caption{Demonstration of the reduction of the number of
measurements as a secondary cost for the water molecule calculation using various optimization schemes.
The calculation is carried out using the HMP2 framework
with the UCCSD ansatz. The total number of expectation values to be evaluated on a quantum computer
as a function of the number $N$ of $D_{\alpha}$ terms in the UCCSD
ansatz operator for a VQE cycle is shown in (a) for
three different optimization schemes. The black solid
line represents the reduction using only QSR. The
blue dotted line shows the reduction using the QWC partition and
QSR at the same time. The red dashed line illustrates the
most reduction, using the GC partition along with QSR.
Note the plateau-like behaviors that make the lines to appear like staircases are due to our QSR technique (orange, dot-dashed line with the $y$ axis on the right), i.e., the jumps in the number of expectation values to be measured per VQE iteration occur when the QSR technique makes a transition from one space requirement to another.
The average number of additional two-qubit gates incurred by using
the GC partition along with QSR, $n_{\mathrm{GC+QSR}}$, 
is shown in (b) as the the blue dotted
line. This may be compared with the black solid line, 
representing the total number of two-qubit gates $n_{\mathrm{JW}}$
required to induce UCCSD ansatz states, implemented with 
the JW transformation.
The distribution of $n_{\mathrm{GC+QSR}}$ is given in the SM Sec.~\ref{app:extra-cnot-distribution}.
The red dashed line with an alternative $y$-axis label
on the right shows the ratio $R = n_{\mathrm{GC+QSR}}/(n_{\mathrm{JW}} + n_{\mathrm{GC+QSR}})$.
}
\label{fig:perturb-observation}
\end{figure}

Figure~\ref{fig:convg} shows the convergence of the ground state energies
using different approaches as the number $N$ of $D_{\alpha}$ terms included in the 
UCCSD ansatz operator increases. 
For the conventional approach \cite{ChemDemon4}, the insertion order
of the ansatz terms is obtained from the prior knowledge of the order of
contribution of determinants in a classical FCI calculation, 
which closely mimics the ideal case but is rarely realistic to obtain.
For the UCCSD energies obtained using the proposed HMP2,
we bootstrapped the ordering of the ansatz terms as detailed in Sec.~\ref{sssec:prtb_hmp2}. 
Comparing the convergence of the UCCSD energies, we find that the
HMP2-bootstrapped ordering effectively captures the major ansatz
terms. This is confirmed by the good agreement between the FCI ordering
and the HMP2 ordering shown in Fig.~\ref{fig:convg}.
We also observe that the HMP2 correction to the ground-state energy 
helps accelerate the energy convergence towards the FCI energy
significantly. Provided that implementation of each additional ansatz term 
leads to a substantial accumulation of noise in the NISQ hardware,
we find the rapid energy convergence enabled by the HMP2 method 
to be particularly useful for the near-term quantum computers.

In addition to focusing on reducing the circuit complexity, 
which is the most important cost for NISQ era applications, 
we examine the total number of measurements as 
a secondary cost using the water molecule 
calculation as an example.
Figure \ref{fig:perturb-observation} (a) shows the number of measurements using various optimization
schemes discussed in Sec.\ref{sssec:prtb_hmp2}. 
Comparing to the number of measurements optimized using only QSR, the number optimized using the
GC partition method and QSR is more than two orders of magnitude smaller while the number is about one order of magnitude smaller using the QWC partition method along with QSR. 
Figure \ref{fig:perturb-observation} (b) shows the additional
cost in terms of the number of two-qubit gates
associated with using the GC partition method along with QSR comparing to 
the total number of two-qubit gates using 
the JW transformation.
The ratio between the two quickly reduces as the number
of ansatz terms increases. Thus one should decide whether to use GC or QWC along with QSR after weighing their benefits 
against their cost on a case-by-case basis.

\subsection{Generalized transformations for Fermion to Qubit operator}
\label{subsec:GM}

In this section, we investigate how a variety of fermion to qubit transformations may be used to reduce the quantum resource requirements for the pre-FT fermion simulations. It is important to note that all of the transformations are equivalent and thus the resulting quantum circuits for each transformation implements exactly the same fermion simulation. Therefore, the resource savings we obtain in this section are independent of the accuracy of the simulations, in general. 

Well-known fermion to qubit transformations, such as Jordan-Wigner (JW) \cite{JW} or Bravyi-Kitaev (BK) \cite{BK} transformations, map fermionic creation (annihilation) operators to Pauli strings. However, there exist numerous other transformations available for use. Below, we introduce a generalized transformation (GT) method (see also \cite{GM}), of which the JW and BK transformations are a part. We show that, when used with the PF algorithm as a concrete example for the implementation of the UCCSD ansatz, significant quantum resource savings may be achieved by a suitable choice of the mapping for a given cluster operator input, together with a carefully chosen sequence of heuristic optimization methods.

All transformations in the GT method must respect relations specified in (\ref{anticom}). This may be achieved by considering the following invertible, upper-triangular basis-transformation matrix $\beta$, which transforms the occupation number basis to a GT basis according to
\begin{align}
\beta = \left(\begin{matrix} 
&\beta_{n-1,n-1} \quad &\beta_{n-1,n-2}& \; ... &\beta_{n-1,0} \\ 
&0           \quad &\beta_{n-2,n-2}& \; ... &\beta_{n-2,0} \\ 
&\vdots                  &    \vdots &        &\vdots          \\
&0           \quad &0          & \; ... &\beta_{0,0}
\end{matrix}\right),
\end{align}
where $\beta_{i,j} \in \{0,1\}$ for $i > j$, $\beta_{i,i} = 1$, and $n$ is the number of qubits involved in the transformation. Following closely the notations used in \cite{ar:SRL}, we define the following sets of indices for convenience, which we detail below. We note that all matrix operations are performed in modulo-2 space and the main diagonal elements are excluded when generating these sets.

\begin{itemize}
\item
Update set $U(j)$: elements of this set are the row indices with non-zero entries in column $j$ of the basis-transformation matrix $\beta$.

\item
Parity set $P(j)$:  elements of this set are the column indices with non-zero entries in row $j$ of the matrix  $(\pi\beta^{-1}-\beta^{-1})$, where $\pi_{i,j} = 1$ if $i\leq j$, otherwise 0.

\item
Remainder set $R(j)$: elements of this set are the column indices with non-zero entries in row $j$ of the matrix $\pi\beta^{-1}$.
\end{itemize}

The GT-based creation and annihilation operators are then
\begin{align}
\label{eq:GMca}
a_j^\dagger &\equiv \left[ \sigma_x^{U(j)} \otimes \sigma_x^{j} \otimes \sigma_z^{P(j)} - i \sigma_x^{U(j)} \otimes \sigma_y^{j} \otimes \sigma_z^{R(j)} \right]/2, \nonumber \\
a_j &\equiv \left[ \sigma_x^{U(j)} \otimes \sigma_x^{j} \otimes \sigma_z^{P(j)} + i \sigma_x^{U(j)} \otimes \sigma_y^{j} \otimes \sigma_z^{R(j)} \right]/2,
\end{align}
which can straightforwardly be shown to satisfy (\ref{anticom}). We note that, e.g., the JW transformation is a special case of $\beta = \mathbf{1}$. 

The state of the art implementation of the UCCSD ansatz~\cite{ChemDemon4} relies on the JW transformation and it considers a series of heuristics that were chosen carefully to optimize the resulting quantum circuits. To enable a proper comparison of those with our results then, it is necessary for us to consider a similar series of heuristics as well. Details of the heuristics we consider are provided in the SM Secs.~\ref{app:label}--\ref{app:beta-choice}. Below, we briefly outline the steps in the order of applications. The cost function we consider for our concrete example is the number of two-qubit gates.

The outer-most loop of our approach considers different transformation matrices $\beta$. For a given mapping matrix $\beta$, we execute the following routines to construct and optimize our circuit that implements the UCCSD ansatz. The routines are designed to repeatedly call a suite of dedicated, automated circuit optimization tools, whose technical details may be found in \cite{ChemDemon4,Optimizer}. The efficiency of these tools allows us to quickly evaluate the cost function for different cases we consider in each of the subroutines.
\begin{enumerate}[label=\bfseries Routine \arabic*,leftmargin=*,labelindent=1em]

\item
\emph{Fermionic level labeling: }
Unlike in the JW transformation where $U(j)$ is an empty set and $P(j) = R(j)$, in the GT approach, myriads of combinations of sets $U(j)$, $P(j)$, and $R(j)$ are possible. To best take advantage of this, it is critical to carefully select which fermion level is mapped onto which qubit index. Exploring all possible mappings is however computationally prohibitively expensive. We thus resort to a simple greedy approach, whereby we explore one permutation at a time from a given fermion-level to qubit-index mapping. Specifically, from the given mapping, we apply the permutation that results in the most reduction in the quantum resource requirements. We iterate this process until no single permutation results in the reduction of quantum resource requirements. See SM Sec.~\ref{app:label} for detail. See \ref{subroutine2} below for the cost function evaluation for each permutation.

\item
\emph{Inter-Trotter term ordering: }
Demonstrated in \cite{poulin2014trotter,hastings2014improving,Wecker_2014,Babbush_2015,Tranter_2019,ChemDemon4} was that ordering the Trotter terms appropriately can lead to large savings in the two-qubit gate counts due to gate cancellations between the neighboring Trotter term circuits. While a similar approach is indeed possible -- and we base our method on the work reported in \cite{ChemDemon4} -- in our GT approach, a non-trivial modification to a method reported in \cite{ChemDemon4} needs to be made. Specifically, we need to preprocess each Trotter term to determine its eligibility for being classified under the same equivalence class, the elements of which have the opportunities for resource savings when placed next to one another on a quantum circuit via the aforementioned gate cancellations. The eligibility criteria are straightforward for the JW transformation, considered in \cite{ChemDemon4}. See SM Sec.~\ref{app:inter-trotter} for details for the GT method. Once the equivalence classes according to the eligibility of each Trotter term is determined, we use a simple greedy approach to order the Trotter-term elements of each equivalence class to reduce the two-qubit gate counts. 

\end{enumerate}
\begin{enumerate}[label=\bfseries Subroutine \arabic*,leftmargin=*,labelindent=1em]

\item
\label{subroutine2}
\emph{Intra-Trotter term ordering: }
For a given fermion-label to qubit-index mapping, an efficient method to implement a single- or a double-fermion excitation Trotter term is known \cite{Whitfield} in the case where the JW transformation is used, and an aggressively optimized circuit construction that leverages full qubit-to-qubit connectivity is available in \cite{ChemDemon4}. The method in \cite{ChemDemon4} relies on a careful ordering of the intra-Trotter operator implementation, where, for instance, an example double-fermion Trotter term $e^{t_{pqrs}(a_p^\dagger a_q^\dagger a_r a_s - h.c)}$ is expanded to  $\sigma_{x,y,z}$-based intra-Trotter terms.
To enable an efficient implementation in other transformations used in our GT approach, we compute the cost function for every possible permutation of the intra-Trotter terms (see SM Sec.~\ref{app:subroutine:intra} for detail). We choose an ordering with the least cost.
\end{enumerate}
The resulting optimized circuit implements the UCCSD ansatz in the chosen transformation basis defined by $\beta$. However, with the exception of the JW transformation where $\beta = {\bf 1}$, the GT $\beta$ matrix requires us to implement the initial mapping of the basis at the beginning of the circuit. This incurs an overhead $O(n^2/\log(n))$ in the two-qubit gate counts~\cite{Igor}. To obtain the final quantum resource requirement, we call an automated optimizer with the input quantum circuit that consists of the prefix subcircuit that implements $\beta$ and the postfix subcircuit that implements the $\beta$-basis UCCSD ansatz. 

We note that in \cite{ChemDemon4} the concept of ``bosonic'' excitations is discussed. Effectively, in the JW transformation, whenever a pair of neighboring qubits, whichever appropriate fermion levels they correspond to, are excited to yet another pair of neighboring qubits that denote another set of fermion levels, the circuit that implements such an excitation term can be dramatically simplified to require only two two-qubit gates, while requiring only half the number of qubits that would otherwise be required (see SM Sec.~\ref{app:GenBos} for the cases where the pairs do not neighbor). To take advantage of this, we use a juxtaposition of the bosonic circuit written according to the JW transformation and non-bosonic circuit written according to our GT approach. We note in passing that, to return from the half-qubit space of the bosonic circuit to the full-qubit space of the non-bosonic circuit, at most $n/2$ $\cnotgate$\ gates are expended. All our circuit metrics appropriately reflect this.

\tab{gm} shows circuit metrics, measured according to the number of two-qubit gates used to implement the UCCSD ansatz circuit, for different molecules of our choice. We show the results for the JW, BK, and the best GT transformations that our heuristic toolchain specified above found for comparison. To find the best GT transformations, we used a particle swarm optimization, as detailed in SM Secs.~\ref{routine:beta} and \ref{app:beta-choice}. The advantages offered by the GT transformations vary in the suite of molecules we consider, ranging from $1.44\%$ to $21.43\%$. This demonstrates the capability of our heuristics that it is indeed possible to further optimize the quantum circuits over the previous state of the art obtained via the JW transformation by considering GT transformations, custom selected for different input cases. 

We make an open-ended note here on the savings ratio obtained. The savings ratio for the GT method is obtained by comparison to the JW transformation, which often admits better circuit optimization over the BK transformation in the instances we considered, especially around the interesting regimes where the ansatz circuit is large enough to recover, e.g., chemical accuracy, but is sufficiently small to merit VQE implementations on a near-term, noisy device. In the asymptotic limit, the BK transformation, one of the GTs, is known to incur $O(\log(n))$ overhead, an exponential advantage over the JW transformation that incurs $O(n)$ overhead~\cite{zhu1997algorithm}. Provided that, in the small instances, the savings ratio can be larger than $20\%$, a question arises as to how the ratio will behave in the intermediate-sized, aforementioned interesting regime. We mention here a couple possibilities. It may be that the improved circuit efficiency by the GT method for a particular excitation term could be offset by the increased cost of implementing other excitation terms. In this case, the ratio is expected to drop. It could also be that the JW transformation, due to its lack of nimbleness in the definition of transformation to help reduce potentially large cost of adding the excitation terms to the ansatz circuit, incurs a large overhead, thus resulting in the more significant savings ratio. Further research is required to provide better predictions. The low reduction rate observed for larger sized circuits in \tab{gm} may in part be due to the ineffectiveness of the classical optimization strategy and the limited classical computational resources used, which is further discussed in SM Sec.~\ref{app:beta-choice}.

\begin{table}[]
\centering
\begin{tabular}{lllrrrr}
Molecule 		&$n$ 	    &NE 	& JW 		& BK  	&GT		&Improve(\%)\\ \hline \hline
\ce{HF}			&6		&3		&30		    &29	    &25	    &16.67 \\ \hline
\ce{LiH}		&6		&3		&30		    &29	    &25	    &16.67\\ \hline
\ce{BeH2}		&8		&9		&70		    &71	    &60	    &14.29\\ \hline
\ce{NH3}		&14		&52		&485		&607	&478    &1.44\\ \hline
& \\ \hline \hline
\ce{H2O}(4)		&8		&4		&42	    &50	    &33		&21.43	\\ \hline
\ce{H2O}(5)	    &8		&5		&44	    &52	    &35		&20.45	\\ \hline
\ce{H2O}(6)	    &8		&6		&46	    &47	    &37		&19.56	\\ \hline
\ce{H2O}(8)	    &10		&8		&68	    &88	    &63	    &7.35\\ \hline
\ce{H2O}(9)	    &10		&9		&71	    &89	    &66	    &7.04\\ \hline
\ce{H2O}(11)	&10		&11		&93	    &110	&87	    &6.45\\ \hline
\ce{H2O}(12)	&10		&12		&95	    &112	&89	    &6.32\\ \hline
\ce{H2O}(14)	&10		&14		&114	&140	&111    &2.63\\ \hline
\ce{H2O}(16)	&10		&16		&135	&166	&131	&2.96\\ \hline
\ce{H2O}(17)	&10		&17		&137	&168	&133	 &2.91\\ \hline
\end{tabular}
\caption{Number of two-qubit gates required for the VQE simulation of different molecules with different fermion to qubit transformations. 
$n$ is the dimension of the $\beta$ matrix. NE is the number of excitation terms considered in the UCCSD ansatz. JW/BK are the number of two-qubit gates with JW or BK transformations. GT is the number of two-qubit gates given by the best $\beta$ other than JW or BK. 
All molecules use the STO-3G basis, and ansatz terms are determined according to the HMP2 ordering. 
(Top): Here, we report the results for the cases where the HMP2 method was used to reach chemical accuracy.
(Bottom): Here, we report the results for the HMP2 progression for a water molecule. The number in the parentheses next to \ce{H2O} indicates the total number of excitation terms $(D_{\alpha})$ considered for the UCCSD ansatz. The cutoff of \ce{H2O}(17) is chosen based on the recovery of chemical accuracy (see \fig{convg}).}

\label{tab:gm}
\end{table}

\section{Discussion} 
\label{sec:disc}

So far in this paper, in the optimization of the FT-regime circuits, we have considered efficient implementations of each Trotter term. It should however be noted that a parallel implementation of multiple Trotter terms should also be considered in the $\rzgate$ gate depth reduction of the quantum simulation circuits. Based on our circuit construction detailed in Sec.~\ref{sec:FT}, we propose the following methodology for optimization over the parallel implementation.

As discussed in Sec.~\ref{sec:FT}, the circuit that implements the Trotter term (see \fig{cirq-FT-opt}) consists largely of three parts: an initial $\cnotgate$ gate network that computes a linear function of Boolean input variables, a triply-controlled $\rxgate$ gate, and the inverse of the initial $\cnotgate$ gate network. Denoting the Boolean variables at the input as, in the order from top to bottom, $a$, $b$, $c$, and $d$, the $\cnotgate$ gate network outputs $a$, $b \oplus c$, $a \oplus c$, and $a \oplus d$. The bottom three outputs remain invariant over the action of the triply-controlled $\rxgate$ gate. This means that any linear functions of $b \oplus c$, $a \oplus c$, and $a \oplus d$ are accessible for use in the implementation of $\cnotgate$\ gates that corresponds to the JW $\sigma_z$ strings for other Trotter terms. This is so, because the invariance is required to implement the appropriate inverse of the {$\cnotgate$\ gates} that were used to take the JW $\sigma_z$ strings into consideration in the first place, as per our circuit construction shown in SM~\fig{cirq-with-z}. Heuristic methods that collect those Trotter terms and simultaneously implement them can then be used to optimize the depth of the quantum circuit.

For the pre-FT regime VQE simulations, we have proposed a general
framework that leverages the predictive and corrective power
of perturbation theory. 
The predictive feature of our framework resembles that
of the ADAPT-VQE method~\cite{ADAPT-VQE}, which aims to iteratively add terms to a generalized Trotterized exponential ansatz by following the steepest descent based on the calculations of numerical gradients. 
Despite trying to solve the same problem, our framework adopts a different philosophy and computes an approximate (perturbative) correction to the wavefunction and chooses the following operator based on how much of a correction an operator could potentially provide. 
Due to the two completely different approaches to the problem, the two methods could have vastly different use cases for which they are suitable respectively. 
To identify the difference in use cases, further thorough investigation into different areas of applications are needed which is beyond the scope of this paper. 

Additionally, our framework is exceedingly flexible and provides means
to balance between resource constraints and accuracy requirement. 
First of all, our framework not only works with the UCCSD ansatz, 
but any unitary ansatz, including the generalized Trotterized exponential 
ansatz used in the ADAPT-VQE method. Second, the predictive feature
and the corrective feature can be used independently from each other. 
For instance, the perturbative corrections can be used along with the 
ADAPT-VQE method, which provides the ansatz term prediction. 
Note that doing so, according to our framework, does not 
incur additional cost in terms of circuit complexity; 
it incurs overhead only in terms of the number of measurements. 
Our specific examples described in this manuscript results in
smaller number of measurements required, since both the
prediction and the correction are obtained by the same set of
measurements, as opposed to two different set of measurements
that would otherwise be required, had we combined ADAPT-VQE 
with our perturbative correction.
Third, the choice of the set of orbital sequence $\{\alpha'\}$ in 
(\ref{eqn:2nd-perturb}) can be leveraged to balance between accuracy
requirement and resource constraints. On one hand, the size of the 
set can be decreased to reduce the number of measurements needed.
On the other hand, it is also possible to include orbital sequences 
outside of the set $\{\alpha\}$ for the ansatz operator in order
to achieve better accuracy without sacrificing circuit complexity.
For instance, one could include triple excitation sequences in $\{\alpha'\}$
along with the UCCSD ansatz to provide a perturbative ``T'' correction
to the final ground state energy. In this case the increased accuracy
does not correspond to increased circuit complexity but only incurs
overhead in terms of the number of measurements.

We note that, extending and generalizing the framework used for considering the bosonic terms in the JW transformation and the non-bosonic terms in the GT transformation, the use of multiple fermion to qubit transformations $\beta$ for a given set of excitation terms could be of value in reducing the overall resource requirement. Drawing from the fact that different $\beta$ transformations result in different resource requirements in implementing the excitation terms of a target UCCSD ansatz circuit, it is reasonable to expect that a certain subset of the excitation terms may be more efficiently implemented by one $\beta$ transformation and some other excitation terms may be implemented more efficiently by yet another $\beta$ transformation. Thus, dividing the set of excitation terms required for preparing a UCCSD ansatz state into subsets of excitation terms that may be more efficiently implemented by respective, appropriate choices of $\beta$ transformations for each of the subsets may prove to be more advantageous in the quantum resource requirement. Optimizing over the tug of war expected between the overhead cost incurred due to the switching of the transformations and the savings obtained via the tailor-made choices of the transformations remains as future work.

\section{Conclusion} 
\label{sec:conc}
Quantum simulations performed by quantum computers have long been thought to be one of the most promising quantum applications that will prove advantageous over classical computers. Despite the recent technological advancements made by the community, there still remains a gap between what a quantum device can realistically achieve and what is required to demonstrate a practical advantage in running a quantum simulation on a quantum device. In an attempt to bridge this gap, in this paper, we focused on optimizing the quantum resource requirements for both the FT and pre-FT regime quantum simulations of fermionic systems. Our FT-regime optimized approach yields a $\rzgate$-depth of one quantum circuit for each Trotterized evolution operator with only two $\rzgate$ gates to implement in total, while requiring just a single ancilla qubit. Our pre-FT regime optimized approach leverages perturbation theory, capable of determining the best ansatz terms to consider to enable a rapid convergence to the sought-after ground-state energy of a simulated molecule that in turn results in the reduction of gate counts required to reach a certain accuracy threshold. We also considered GT methods to obtain significant quantum resource savings, upwards of 20\% in some instances, over the conventional JW or BK transformations in practice. While these results were applied to a specific set of fermionic systems in this paper as a concrete example, we expect similarly-spirited works could leverage our methodologies. We believe the savings in the quantum resource requirement demonstrated by our approaches help bring the day of solving practical problems using a quantum computer closer.

\section*{Acknowledgement}
This work was supported, in part, by the DOE BES QIS Program (de-sc0019449).
The authors thank Diego Acosta and Germ{\'a}n Mauricio Mu\~noz at IonQ, Inc., for their efforts in the implementation of computing systems used to generate the data presented in this manuscript.
The authors further thank Prof. Reinhold Bl\"umel for helpful discussion.

\bibliographystyle{plainnat}
\bibliography{main}% Produces the bibliography via BibTeX.

\onecolumn\newpage
\appendix

\section*{SUPPLEMENTARY MATERIAL}
\section{Circuits for the fault-tolerant regime}
\label{app:FTcirc}

In this section, we show in detail the circuit
implementation of the Trotter terms, typically 
encountered in quantum chemistry simulations,
optimized for the FT regime. First, we show
how to implement a triply-controlled $\rxgate$ gate,
needed for our two-body term circuit shown in
\fig{cirq-FT-opt} of the main text. Then, we compare the
resulting quantum resource requirements of our
method against those of the method shown in \cite{Elucidating},
which can be optimized by combining with the work
discussed in \cite{Gidney_2018,ar:LowCost}, in terms 
of circuit size, width, and depth. 

\fig{cirq-triply}(a) shows an $\rzgate$- and $\tgate$-gate optimized
implementation of a triply-controlled $\rxgate$ gate. 
We use the so-called relative phase Toffoli gate investigated in 
\cite{RTOF} instead of a conventional Toffoli gate to reduce
the $\tgate$ gate count, excluding that required for the
$\rzgate$ gates, to 16. An additional advantage is that
a relative phase Toffoli gate with three controls does not require
an ancilla qubit, whereas a conventional triply-controlled Toffoli 
gate, decomposed into the standard gate set of $\cnotgate$, 
single-qubit Clifford, and $\tgate$ gates, requires at least
one ancilla. For completeness, we show in \fig{cirq-reltof}
the explicit construction of the triply-controlled, relative phase
Toffoli gate, imported verbatim from \cite{RTOF},
although other methods, sometimes with a different gate set,
have been reported elsewhere~\cite{CZgate,GlobalMS}.

Inspecting \fig{cirq-triply}(a), we notice that we need to 
apply the two $\rzgate$ gates in series. To reduce the 
$\rzgate$ depth then, we can introduce an ancilla qubit
prepared to $\ket{0}$. An $\rzgate$-depth one construction
of the triply-controlled $\rxgate$ gate is shown in 
\fig{cirq-triply}(b). Should the same angle rotations
happen in multiple qubits at the same time, a possibility
if a molecule of interest that is being simulated has
symmetries, a weight-sum trick reported and used in
\cite{Gidney_2018,ar:LowCost} can be employed to exponentially
reduce the number of $\rzgate$ gates at a modest cost
in the number of qubits and $\tgate$ gates.
Two $\rzgate$ gates to be applied in parallel 
however do not benefit from this trick.

\begin{figure}
\[
(a)\hspace{3mm}
\Qcircuit @C=0.4em @!R {
&\ctrl{1}                       & \qw \\
&\ctrl{1}                       & \qw \\
&\ctrl{1}                       & \qw \\
&\gate{\rxgate(\theta)}& \qw
}
\hspace{3mm}\raisebox{-21.0mm}{=}\hspace{3mm}
\Qcircuit @C=0.4em @!R {
&\qw                &\ctrl{1}^{R^{ }} &\qw                                  &\ctrl{1}^{R^{-1}} &\qw                                    & \qw               & \qw \\
&\qw                &\ctrl{1}            &\qw                                   &\ctrl{1}               &\qw                                   &\qw                & \qw \\
&\qw                &\ctrl{1}            &\qw                                   &\ctrl{1}               &\qw                                   &\qw                & \qw \\
& \gate{\hgate} &\targ               &\gate{\rzgate{(-\theta/2})} &\targ                  &\gate{\rzgate{(\theta/2})}  &\gate{\hgate} & \qw
}
\]

\vspace{1cm}

\[
(b)\hspace{3mm}
\Qcircuit @C=0.4em @!R {
& \qw& \qw&\ctrl{1}                       & \qw & \qw & \qw \\
& \qw& \qw&\ctrl{1}                       & \qw & \qw & \qw\\
& \qw& \qw&\ctrl{1}                       & \qw & \qw & \qw\\
& \qw& \qw&\gate{\rxgate(\theta)}         & \qw & \qw & \qw\\
\ket{0}&  &  & \qw                       & \qw &     & \ket{0}
}
\hspace{3mm}\raisebox{-29.0mm}{=}\hspace{3mm}
\Qcircuit @C=0.4em @!R {
&\qw                &\ctrl{1}^{R^{ }} &\qw     &\qw                                  &\qw     &\ctrl{1}^{R^{-1}\;\;} & \qw               & \qw \\
&\qw                &\ctrl{1}            &\qw     &\qw                                   &\qw     &\ctrl{1}               &\qw                & \qw \\
&\qw                &\ctrl{2}            &\qw     &\qw                                   &\qw     &\ctrl{2}               &\qw                & \qw \\
&\gate{\hgate} &\qw                &\ctrl{1} &\gate{\rzgate{(\theta/2})}  &\ctrl{1} &\qw                   &\gate{\hgate} & \qw \\
\ket{0}&           &\targ              &\targ   &\gate{\rzgate{(-\theta/2})}  &\targ   &\targ                  &\qw                 &\ket{0}
}
\]
\caption{\label{fig:cirq-triply} Triply-controlled $\rxgate (\theta)$ gate implementation with relative-phase triply-controlled \notgate\ gates. The $R$-prepended triply-controlled Toffoli gate denotes a relative-phase Toffoli gate, i.e., the gate implements a triply-controlled-$\notgate$ gate up to relative phases. The $R^{-1}$-prependix is used to denote its inverse. The method to implement the relative-phase Toffoli gate is detailed in \cite{CZgate,RTOF,GlobalMS}.}
\end{figure}

\begin{figure}
\[
\Qcircuit @C=0.4em @R=1.2em {
	&\qw        &\qw        &\qw        &\qw                &\qw        &\ctrl{3}   &\qw        &\qw        &\qw                &\ctrl{3}   &\qw        &\qw        &\qw                &\qw        &\qw        &\qw        &\qw                &\qw        &\qw\\
	&\qw        &\qw        &\qw        &\qw                &\qw        &\qw        &\qw        &\ctrl{2}   &\qw                &\qw        &\qw        &\ctrl{2}   &\qw                &\qw        &\qw        &\qw        &\qw                &\qw        &\qw\\
	&\qw        &\qw        &\ctrl{1}   &\qw                &\qw        &\qw        &\qw        &\qw        &\qw                &\qw        &\qw        &\qw        &\qw                &\qw        &\qw        &\ctrl{1}   &\qw                &\qw        &\qw    \\
	&\gate{H}   &\gate{T}   &\targ      &\gate{T^\dagger}   &\gate{H}   &\targ      &\gate{T}   &\targ      &\gate{T^\dagger}   &\targ      &\gate{T}   &\targ      &\gate{T^\dagger}   &\gate{H}   &\gate{T}   &\targ      &\gate{T^\dagger}   &\gate{H}   &\qw
}
\]

\caption{\label{fig:cirq-reltof} A quantum circuit for a relative-phase triply-controlled $\notgate$ gate. The circuit is taken verbatim from \cite{RTOF}.
}
\end{figure}

\begin{figure}
\[
\Qcircuit @C=0.4em @R=1.2em {
	+\; & &\qw &\ctrl{1} &\gate{\rxgate(\theta)} &\ctrl{1} &\qw \\
	-\; & &\qw  &\targ    &\ctrl{-1}                      &\targ    &\qw
}
\]
\caption{\label{fig:cirq-single} A circuit construction of a single-body term that is in analogy to the two-body circuit discussed earlier. The circuit is equivalent to that reported in Fig.~16 of \cite{singleref1} or Fig.~9 of \cite{singleref2} up to a basis transformation.
}
\end{figure}

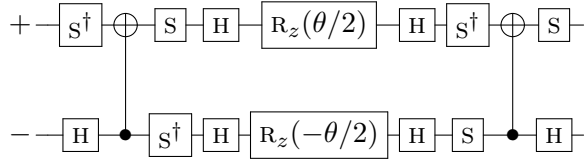
\begin{figure}
\[
\Qcircuit @C=0.4em @!R {
	+\; & &\qw &\gate{\sgate^\dag} &\targ     &\gate{\sgate}      &\gate{\hgate} &\gate{\rzgate(\theta/2)}  &\gate{\hgate} &\gate{\sgate^\dag} &\targ     &\gate{\sgate} &\qw \\
	-\; & &\qw &\gate{\hgate}      &\ctrl{-1} &\gate{\sgate^\dag} &\gate{\hgate} &\gate{\rzgate(-\theta/2)} &\gate{\hgate} &\gate{\sgate}      &\ctrl{-1} &\gate{\hgate} &\qw
}
\]
\caption{\label{fig:cirq-control-rx} A depth-optimized construction of a single-body term in Fig.~\ref{fig:cirq-single}.
}
\end{figure} 

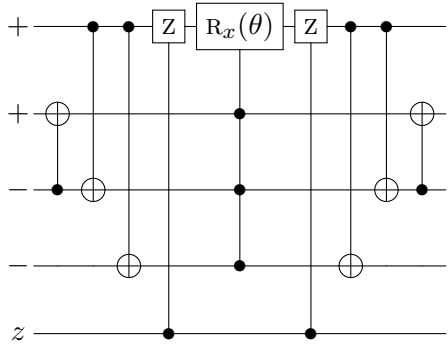
\begin{figure}
\[
\Qcircuit @C=0.4em @R=1.8em {
	+\; & &\qw       &\ctrl{2} &\ctrl{3} &\gate{\zgate} &\gate{\rxgate(\theta)} &\gate{\zgate} &\ctrl{3} &\ctrl{2} &\qw        &\qw \\
	+\; & &\targ     &\qw      &\qw      &\qw       &\ctrl{-1}                      &\qw       &\qw     &\qw      &\targ      &\qw \\
	-\; & &\ctrl{-1} &\targ     &\qw      &\qw       &\ctrl{-1}                      &\qw       &\qw     &\targ     &\ctrl{-1} &\qw \\
	-\; & &\qw       &\qw       &\targ    &\qw       &\ctrl{-1}                      &\qw       &\targ   &\qw       &\qw      &\qw \\
	z\; & &\qw       &\qw      &\qw      &\ctrl{-4} &\qw                            &\ctrl{-4} &\qw     &\qw       &\qw      &\qw
}
\]

\caption{\label{fig:cirq-with-z} FT-regime optimized circuit for the two-body term $e^{-i\theta/2(\sigma_{+}\sigma_{+}\sigma_{-}\sigma_{-}\sigma_z + {\rm h.c.})}$. We marked each qubit lines with $+$, $-$, or $z$ to denote which qubits are associated with $\sigma_+$, $\sigma_-$, or $\sigma_z$, respectively, for this particular example.}
\end{figure}

Having fully described our circuit that implements a
two-body term, we are now ready to compare it with
the state of the art in \cite{Elucidating}. 
Figure~6 of the appendix of \cite{Elucidating}
shows a circuit that implements a two-body term,
albeit with different coefficients for each of the eight Pauli
strings obtained by expanding $\sigma_{\pm}$
with $\sigma_{x,y}$ in a two-body term.
Circuit-width wise, four ancilla qubits are used.
$\rzgate$ depth is one.

Since a two-body term of our interest does not incur
different coefficients, the circuit shown in Figure~6
of the appendix of \cite{Elucidating} can be further
optimized using the aforementioned weight-sum trick
in \cite{Gidney_2018,ar:LowCost}. Using various
formulas reported in \cite{ar:LowCost},
we arrive for our purposes at the following
requirements: at most 32 $\tgate$ gates to compute the weight, 
where four full and three half adders are used,
and seven additional ancilla qubits for each adder. 
The number of $\rzgate$ gates to be applied is then reduced to four.
In total, the optimized circuit requires
32 $\tgate$ gates, 4 $\rzgate$ gates in one layer,
and eleven ancilla qubits, four from Fig.~6 of the appendix
of \cite{Elucidating} and seven from the weight-sum trick
reported in \cite{Gidney_2018,ar:LowCost}.
This may be compared with our $\rzgate$-depth optimal
implementation that requires 16 $\tgate$ gates, two
$\rzgate$ gates in one layer, and one ancilla qubit.

We next briefly describe a single-body term, reported elsewhere \cite{singleref1,singleref2} before, for completeness. Our work reported above may be considered as a generalization of it to a multi-body term. Note, while we do not explicitly show an $N$-body term circuit as we do not need it for our purposes, an optimal construction can straightforwardly be derived using the work reported herein.

For the single-body operator $e^{-i\theta/2(\sigma_+\sigma_- + {\rm h.c.})}$, in analogy to the two-body case shown above, we may use the circuit shown in \fig{cirq-single} to implement the operator. An $\rzgate$-depth optimized version is shown in \fig{cirq-control-rx}. This results in a circuit that requires no ancilla qubit and two $\rzgate$ gates in one layer.   

We note in passing that if there are $\sigma_{x,y,z}$ operators that act on some other qubits, e.g., due to the JW transformation, we may simply modify the two-body circuit as follows. Consider, for example, there is a $\sigma_z$ operator acting on an additional qubit. In this case, to implement $e^{-i\theta/2(\sigma_{+}\sigma_{+}\sigma_{-}\sigma_{-}\sigma_z + {\rm h.c.})}$, we can modify the circuit in \fig{cirq-FT-opt} and obtain the circuit shown in \fig{cirq-with-z}. A similar treatment is straightforwardly applicable for an arbitrary many body term.

\section{Fermionic level labeling}
\label{app:label}

In this section, we detail the heuristic described in 
{\bf Routine 1} in Sec.~\ref{subsec:GM} which aims to
reduce the number of two-qubit gates by exploiting
different mappings from labels of fermion levels to qubit indices.
Note that the Pauli strings for an excitation operator, after applying the fermion ${\mapsto}$ qubit transformation of choice, depends on the qubit index values according to (\ref{eq:GMca}). 
A better optimized mapping from labels of fermion levels to qubit indices results in fewer Pauli matrices in a given Pauli string which leads to a smaller number of two-qubit gates.
Notice that such a mapping is equivalent to maintaining a
simple mapping from any fermion level $k$ to qubit index $k$
while rearranging the labeling of the fermion levels 
which is entirely arbitrary up until this point.
Thus we may use the freedom of fermion level labeling to reduce the number of two-qubit gates required to implement the one- and two-body Trotter terms.
For an $n$-qubit system representing $n$ fermion levels,
exploring all possible permutations of fermion level labeling is not computationally scalable as the total number of permutations is $n!$. We thus resort to a simple greedy approach as follows.
\begin{enumerate}
\item We first generate a set of unique permutation matrices $\{P_i\}$ for all possible $k$-swap operations.
Here we define a $k$-swap operation as swapping $k$ unique ordered pairs of indices $(2l_j, 2l_j+1)$ with $k$ other unique ordered pairs of indices $(2m_j, 2m_j+1)$ with the same subscript index
$j\in \{0,\cdots,k-1\}$ where $l_j \in \{0,\cdots,n/2-2\}$ and $m_j\in\{l_j+1,\cdots,n/2-1\}$.
We use $k=2$ for all of the examples in the main text.

\item
Denoting the initial fermion labels as a column vector $L_0$, we can then go through all the permutation matrices in the first round and compute the number of two-qubit gates that corresponds to each instance of relabeled fermion levels $P_iL_0$ by simply adding the number of two-qubit gates required for each excitation term obtained from the intra-Trotter term reordering subroutine.
When going through the permutation matrices,
if any matrix results in a lower two-qubit gate
count than all the previous matrices, we record that
two-qubit gate count as $N_1$ and the corresponding 
relabeled level vector as $L_1$. 

\item  
We proceed with the greedy approach iteratively following the first round.
Denoting the resulting relabeled fermion level vector from the $m$th round as $L_m$ and the corresponding two-qubit gate count as $N_m$, 
we again go through all the permutation
matrices and compute the corresponding two-qubit gate
counts for each instance of relabeled levels $P_iL_m$.
Taking $N_{m+1} = N_m$ initially, every time a matrix
results in a two-qubit count lower than $N_{m+1}$, 
we record that gate count as $N_{m+1}$ and the matrix
as $L_{m+1}$.
If there 
is not any matrix that results in a lower two-qubit count 
than $N_m$, we terminate the iteration and 
return $L_m$ as the optimal labeling.
Otherwise we proceed to the next round.
\end{enumerate}
We note that once the labels are determined, they are applied to all relevant fermionic operators, including the molecular Hamiltonian, to be consistent throughout our simulation.

\section{Intra-Trotter term ordering}  
\label{app:subroutine:intra}

In this section, we provide the details of {\bf{Subroutine 1}} described in Sec.~\ref{subsec:GM} which aims to reduce the number of two-qubit gates by optimizing the ordering of the Pauli strings transformed from a two-body operator.

We start by briefly noting that each one-body term leads to only one ordering, up to inversion. This is so, because it contains only two Pauli strings. Therefore, the one-body term does not require a specific ordering.

We next consider two-body operators. A single two-body operator, after applying proper transformation and PF algorithm, contains eight subterms. Specifically, we have
\begin{equation}
U^{{\rm two-body}}  = \prod_{j=0}^7 e^{{-i\theta \otimes_v {\sigma}^{(j,v)}/2 }},
\end{equation}
where $j$ denotes the intra-Trotter term index, $v$ denotes the qubit index, and ${\sigma} \in \{\mathbf{1}, \sigma_x, \sigma_y, \sigma_z \}$.
Each of the intra-Trotter terms $e^{{-i\theta \otimes_v {\sigma}^{(j,v)}/2}}$ can be readily translated into a standard circuit as described in Eq.~(8) in Sec.~Methods~H of \cite{ChemDemon4}. 
In a naive implementation of $U$ that uses an arbitrary fermion to qubit transformation, each intra-Trotter term results in $2(N_j-1)$ number of $\cnotgate$ gates, where $N_j$ is the number of non-identity  ${\sigma}^{(j,v)}$ in the $j$th Pauli string.

Compared to the JW transformation, in the GT method, it is useful to have a more concrete set of rules to determine which qubit may be used as a target qubit $t$, in relation to the realization of $U$ as a quantum circuit (see, e.g., Eq. 8 (b) in Sec.~Methods~H of \cite{ChemDemon4}, for the JW transformation). In the JW transformation, any choice of $t$ such that ${\sigma}^{(j,t)}$ are either $\sigma_x$ or $\sigma_y$, i.e., the qubits that correspond to the fermion labels in the two-body term, are good choices. However, if ${\sigma}^{(j,t)}=\mathbf{1}$ in any one of the eight Pauli strings in $U$ above, which frequently occur in the GT method, the intended target qubit index $t$ is unusable. 
Therefore, in the GT method, qubit choices $t$ that lead to ${\sigma}^{(j,t)}=\mathbf{1}$ are removed from the set of eligible targets that start with all qubits that correspond to the fermion labels that appear in the two-body term of interest and will subsequently not be used in the inter-Trotter term reordering subroutine detailed in Sec.~\ref{app:inter-trotter}. 

\begin{figure}[H]
\[
  \Qcircuit @C=0.7em @R=.2em @!R {
  \vdots & & & & & &  \cdots \\
v\ \ \ \  &\qw & \gate{M_0}		&\ctrl{2}	&\qw				&\ctrl{2}	& \gate{M_0^\dagger} 	 & \gate{M_2}	&\ctrl{2}	&\qw					&\ctrl{2}	& \gate{M_2^\dagger}  	 &\qw  \\
\vdots & & & & & & \cdots \\
t\ \ \ \ &\qw & \gate{M_1}  	&\targ	&\gate{R_z(\theta)}	&\targ	& \gate{M_1^\dagger}     & \gate{M_3}  	&\targ	&\gate{R_z(\theta')}	&\targ	& \gate{M_3^\dagger} 	&\qw \\
\vdots & & & & & &  \cdots
}
\]
\caption{Example circuit for adjacent Pauli string circuits $e^{-i \frac{\theta}{2} {\sigma}_0^v \otimes {\sigma}_1^t \otimes \cdots} e^{-i \frac{\theta'}{2} {\sigma}_2^v \otimes {\sigma}_3^t \otimes \cdots  }$. Subscripts $l$ to Pauli operators are introduced to conveniently label the associated operators $M_l$ shown in the figure. $t$ denotes the target qubit, $v$ denotes the control qubit, ${\sigma}_l \in \{ \sigma_x, \sigma_y, \sigma_z \}$, and  $M_l \in \{ \hgate,\sgate^\dagger \hgate, \mathbf{1}\}$. If ${\sigma}_l = \sigma_x$, $M_l = \hgate$. If ${\sigma}_l = \sigma_y$, $M_l = \sgate^\dagger \hgate$. If ${\sigma}_l = \sigma_z$, $M_l = \mathbf{1}$.} 
\label{fig:Trotter-reorder}
\end{figure}
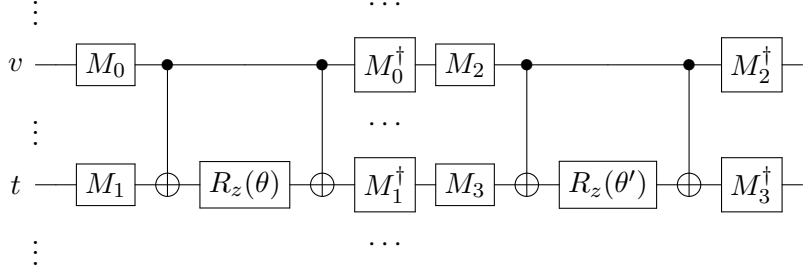

 Now, for a chosen target qubit of choice $t$, we consider the ordering of intra-Trotter terms to maximize the \cnotgate\ reduction. In particular, we only consider the \cnotgate\  reduction between adjacent Pauli string circuits. 
 For each of the adjacent intra-Trotter terms, i.e., $e^{-i\frac{\theta}{2}\otimes_v {\sigma}^{(j,v)}}$ and $e^{-i\frac{\theta'}{2} \otimes_v {\sigma}^{(j+1,v)} }$, 
 where $j\in[0,6]$, we enumerate $v$ through all non-target qubits. We then compare ${\sigma}^{(j,v)}$ and ${\sigma}^{(j+1,v)}$. 
 For a particular control qubit $v$, if neither of the two aforementioned Pauli matrices ${\sigma}^{(j,v)}$ and ${\sigma}^{(j+1,v)}$ is $\mathbf{1}$, then the circuit can be expressed as \fig{Trotter-reorder}. In this particular circuit, if we have $M_0 = M_2$, there is a two \cnotgate\ reduction, while, if we have $M_0 \neq M_2$, there is a one \cnotgate\ reduction, according to the circuit identity discussed in \cite{ChemDemon4} Sec. Method H. Suppose now there are $m_j^{(t)}$ two-\cnotgate\ reductions and $n_j^{(t)}$ one-\cnotgate\ reductions for the chosen target qubit $t$. Then, the total number of \cnotgate\ reductions is $2m_j^{(t)}+n_j^{(t)}$ for the target qubit of choice $t$. 
The total optimized number of \cnotgates\ for a chosen target $t$ of a certain ordering is thus $N^{(t)} = \sum_{j=0}^7 2(N_j-1)-\sum_{j=0}^6 (2m_j^{(t)}+n_j^{(t)})$. 
 
At this point, we are equipped to compute and compare $N^{(t)}$ among all possible orderings and all possible targets to determine which one of the target qubit sets and the intra-Trotter orderings result in the minimum number of \cnotgates. 
We note that for a certain ordering, there likely is a degeneracy in the choice of target qubit $t$ that results in the same \cnotgate\ counts. All of these degenerate cases with the optimal resources are returned to the Inter-Trotter term ordering routine, detailed in Sec.~\ref{app:inter-trotter}, for use.

\section{Inter-Trotter term ordering}
\label{app:inter-trotter}

In this subsection, 
we detail a heuristic to reduce the number of two-qubit gates by taking advantage of the freedom to arbitrarily order the Trotter terms. This is justified since any orderings respect the error bound provided by the PF algorithm.
Denoted as {\bf{Routine 2}} in Sec. \ref{subsec:GM}, this subroutine requires us to first run a preprocessing step based on the information passed from the intra-Trotter term optimization step described in Sec.~\ref{app:subroutine:intra}. Specifically, for operators $a^\dagger_i a_j$ and $a^\dagger_i a^\dagger_j a_k a_l$, we check if any one of the indices $i$ and $j$ for the former, and $i,j,k,$ and $l$ for the latter, may or may not be used as a target qubit in the circuit implementation of the operators in the standard compilation (see, e.g., Fig. 2d of ~\cite{ChemDemon4} for a two-body term using the JW transformation). 
See Sec.~\ref{app:subroutine:intra} for details. We flag the qubit indices that cannot be used as a target as ineligible.

After all of the ineligibilities have been determined, we proceed according to a simple greedy approach. We first identify the most frequently eligible index, say $p$, across all terms. Then, we group all terms with the index $p$ as an eligible target qubit and classify them under the equivalence class $[p]$. We next remove all the group elements from the list of one- and two-body operators. We repeat the procedure from the identification of the most frequent eligible index until no more operators are left in the list.

Note that the quantum resource cost reduction will likely now result in between the circuit representation of the elements of the same equivalence class, since the target qubit of two-qubit gates from each element of the same class is the same. Therefore, once all of the equivalence classes are specified, we consider permuting the orderings by which the elements are implemented on a quantum circuit. Considering all permutations can be prohibitively expensive. We thus use a simple greedy approach once more, first starting out with two of the elements that result in the most resource cost reduction. Then, we concatenate a next element, identified from the set of elements that have not been implemented in the circuit, based on the resource cost reduction. We repeat the concatenation process until no more element is left in the set. We note that each trial 
of testing out which element may be the best for the given iteration consists in general of four cases. This is so, since the circuit concatenation may be performed as a prefix or suffix, and the element to be concatenated can be considered in its original intra-term order or the reverse.

\section{Generalized Bosonic terms}
\label{app:GenBos}

 The correspondence between Generalized Bosonic terms and Fermionic terms can be established as follows.
 A Fermionic double excitation term $\theta_{pqrs} (a_p^\dagger a_q^\dagger a_r a_s-h.c.)$, when transformed by the JW transformation, turns into $\theta_{pqrs} ( \sigma_+^p \sigma_+^q \sigma_-^r \sigma_-^s \otimes_k \sigma_z^k - h.c.)$.
 If $p$ and $q$ belong to the same spatial orbital and $r$ and $s$ belong to yet another same spatial orbital, assuming no other terms that break the symmetry between $p$ and $q$ or $r$ and $s$ have been considered in the circuit, $p$ and $q$ levels may be encoded by a single qubit and likewise for the $r$ and $s$ levels. In this case, using $p$ and $r$ as representatives, the qubit-space operator can be simplified to 
  $\theta_{pqrs} ( \sigma_+^p  \sigma_-^r  \otimes_{k'} \sigma_z^{k'} - h.c.)$, where $k'$ runs over the set of qubits $\sigma_z$ operator needs to be applied for the given excitation term in the appropriately reduced space. The appropriately reduced space may include the single qubits that each denotes the reduced, symmetric levels and those that are not reduced. To illustrate, if two of the levels $k_1$ and $k_2$ in the original space require $\sigma_z$ and if $k_1$ and $k_2$ are encoded into a single index $k'$, we simply call $\sigma_Z^{k'}$ twice, one each for $k_1$ and $k_2$. This for instance amounts to identity, which results in resource savings.

\section{Binary particle swarm optimization} 
\label{routine:beta}

In this section we detail the procedure for using binary PSO~\cite{bpso} to optimize the binary matrix $\beta$ for GT.
To encode the problem, we map the upper triangular entries in an
$n\times n$ binary matrix $\beta$ to a one dimensional binary vector $X\in \{0,1\}^{d} $ where size $d=n(n-1)/2$. Vector $X$
then serves as the location vector for PSO.
The cost function for PSO at a location $X$ mapped from a $\beta$ matrix is then defined as the number of two-qubit gates needed to implement a UCCSD evolution operator using GT with the $\beta$ matrix and subsequently a series of heuristics described in Sec.~\ref{subsec:GM}.

We start by creating a swarm of particles, each with an
initial location $X^i(t = 0)$ mapped from a $\beta$ matrix
sampled from the set of upper triangular matrices 
whose diagonal elements are ones and off-diagonal 
elements are zeros except for $k$ of them. 
The non-negative integer $t$ represents the time step
and $t = 0$ is the beginning of the optimization.
The total number of particles in the swarm is
$\sum_{j=1}^{k}\binom{d}{j}$
where $\binom{\;\sbullet[0.4]\;}{\;\sbullet[0.4]\;}$ denotes a 
binomial coefficient.
Specifically for our examples, we vary $k$ from $1$ to
$k_{\rm max}$ and we have $k_{\rm max} = 6$ for $n \le 8$ and
$k_{\rm max} = 3$ for the rest of the cases.
Each particle also has a velocity vector $V^i(t)$
of the same dimension as $X^i(t)$. The initial matrix
elements of the velocity vectors are taken to be zero.

At every time step $t$, we also keep track of a local optimal
location vector $L^i(t)$ and the corresponding local optimal
cost $s^i(t)$ for each particle $i$ and a global 
optimal location vector $G(t)$.
The local optimal location vector $L^i(t)$ is defined as 
the location vector with the lowest cost function value $s^i(t)$
across all the locations particle $i$ has traveled to 
up to the time step $t$. The global location vector
is the $L^i(t)$ with the lowest $s^i(t)$ among all 
the particles.
The velocity vectors are updated from the $t$th step to the $(t+1)$th
step according to
\begin{align}
    V^i_j(t+1) = w V^i_j(t)  + c_1 (L^i_j(t) - X^i_j(t)) + c_2(G_{j}(t) - X^i_j(t)) \;,
\end{align}
where $w$ is the inertia parameter, $c_1$ is the cognitive parameter, and $c_2$ is the social parameter.
For the examples here, we use various parameters
with $w \in [-4, 4]$, $c_1 \in [0, 2]$, and $c_2 \in [0, 2]$.
The subscript $j$ denotes the $j$th element of the corresponding
vector.
The elements of the location vectors $X^i_j(t+1)$ are 
then updated to $0$ if ${\rm rand}() \le 1/[1+e^{-V^i_j(t+1)}]$
or $1$ otherwise. Here ${\rm rand}()$ is a real quasirandom number
generating function that samples from the uniform distribution on $[0.0,1.0]$.

In practice, binary PSO is usually terminated when
$t$ reaches a predetermined maximal time $t_{\rm max}$.
A particular particle is also stopped when its 
location vector oscillates between two
vectors for more than $\delta t_{\rm osc}$ times.
To further save computation time spent on binary PSO,
we impose a third rule that we stop a particle if
its location vector maintains a Hamming distance larger 
than $s$ from its initial location without reaching a lower
two-qubit gate counts for more than $\delta t_s$ steps.  
Such a termination condition allows similar search space sizes
for $\beta$ matrices of different sizes and optimization
runs with different initial conditions.
For our examples, we use $\delta t_{\rm osc} = 10$,
$s = 6$, $\delta t_s = 10$, $t_{\rm max} = 10000$ for $n \le 8$,
and $t_{\rm max} = 100$ for the rest of the cases.
In most runs in our examples, our choice of $t_{\rm max}$ 
is large enough that the optimization starts to oscillate 
before $t$ reaches $t_{\rm max}$.

\section{General
transformation results using binary particle swarm optimization}
\label{app:beta-choice}
 
In this section we present and discuss examples of GT results 
using binary PSO with various amounts of classical computing resources.
Specifically, we are interested in the fractional improvement in two-qubit gate count, $\rho$, defined according to
\begin{align}
    \rho = \frac{f_{\text{GT}}(R)}{f_{\text{GT}}(R)-f_{\text{JW}}} \;,
\end{align}
where $f_{\text{GT}}$ and $f_{\text{JW}}$ are the numbers of two-qubit gates needed to implement a UCCSD evolution operator using GT with binary PSO or JW, respectively. 
$f_{\text{GT}}$ depends
on the classical computing resources used by binary PSO
which are determined mostly by the number of particles used
since $s$ and $\delta t_s$ are the same for all optimization runs
and $t_{\rm max}$ is always set to be sufficiently
large for our examples as discussed in SM Sec.~\ref{routine:beta}.
To contrast the amount of classical computing resources
used with what is maximally possibly needed, we define a
fractional classical computing resource metric as the
ratio between the number of particles used, $N$, and
the total number of possible variations of the $\beta$ matrices, which can be written as
\begin{align}
    R =  \frac{N}{2^{n(n-1)/2}},
\end{align}
where $n$ is the number of diagonal elements of the $\beta$ matrices. 
Since the number of particles increase with $k$ as discussed in SM Sec.~\ref{routine:beta}, we change $k$ in our examples to effectively
change the amount of classical computing resources used for binary PSO.

\begin{figure}[htbp]
   \centering
   \includegraphics[width=120mm]{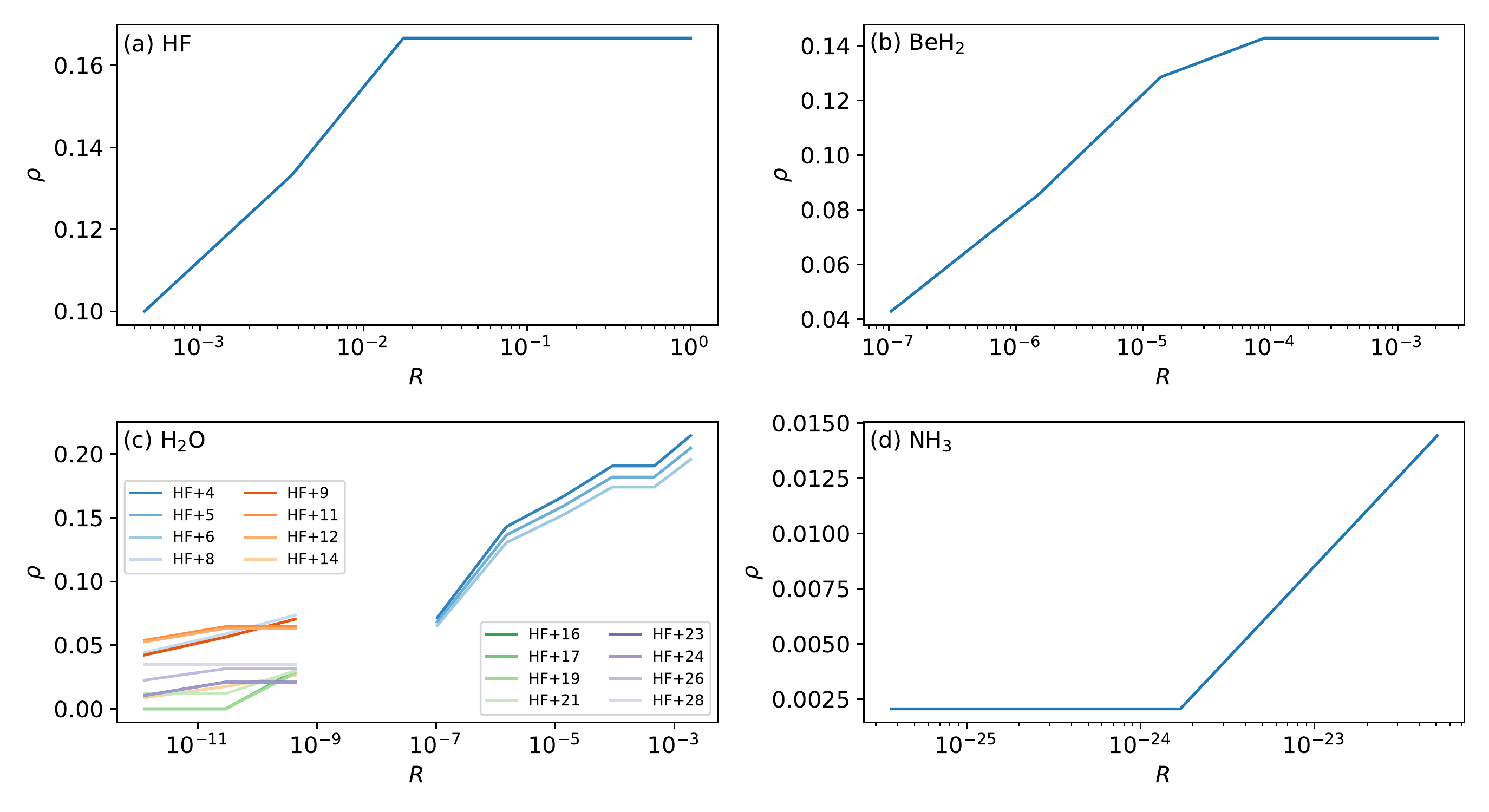} 
   \caption{ 
   Fractional improvement $\rho$ as a function of $R$, the fractional classical computing resources used for binary PSO. (a) \ce{HF}, (b) \ce{BeH2}, (c) \ce{H2O}, and (d) \ce{NH3}.
   All molecules are simulated with the STO-3G basis set using the HMP2 method with UCCSD ansatz.
   For \ce{HF}, \ce{BeH2}, and \ce{NH3}, the number of excitation terms included in the final UCCSD ansatz operators are 3, 9, and 52, respectively, which result in their respective ground state energy estimates within
   chemical accuracy from their known ground state energies.
   The size of the $\beta$ matrices given by their respective
   number of diagonal terms $n$ are shown in Table \ref{tab:gm}.
    }
   \label{fig:improvement}
\end{figure}

Figure~\ref{fig:improvement} shows examples of 
the fractional improvement, $\rho$, as a function of 
the fractional classical computing resources used, $R$, for 
a few different molecules as well as for different numbers of
ansatz terms used in the simulation of the water molecule. 
All the cases are simulated with the STO-3G basis set. 
We observe that $\rho$ increases monotonically with $R$ 
in all cases simulated. 
In some cases, ${\it i.e.}$ hydrogen fluoride (HF), 
beryllium hydride (BeH$_2$), and water molecules with 
four to six ansatz terms included where $n$ is small enough 
that we are able to explore a relatively large portion of
possible variations of $\beta$ matrices ($R > 10^{-3}$), 
we are able to
obtain significant fractional improvement ranging from 
$\approx 14\%$ to over $20\%$.
In other cases where $n$ is large and we are only able
to sample exponentially small portions of the location vector
space in PSO, the fractional improvement obtained is
much smaller and on the single digit percentage level.
Thus we expect, based on the set of numerical results, that more cleverly exploring a large search space is likely a key to increasing the value of $\rho$; Note the cost function $f_{\rm GT}(\cdot)$ landscape is likely complex, making it tricky for binary PSO to avoid being trapped in local minima -- We indeed observed many PSO jobs undergoing oscillatory behaviors.

\section{Qubit space reduction}
\label{app:reduce-qubit-space}

In this section, we detail the procedures for the qubit space reduction (QSR) technique that
optimizes the number of Pauli strings to be measured used in the main text.
There are roughly two categories we consider:
One is to treat the qubits that are in classically accessible states classically
and the other is to take advantage of Bosonic states, i.e.,
we expend only a single qubit for the two spin orbitals in the same spatial 
orbital with degenerate energy due to a particular choice of the ansatz.

As to the former, consider a Pauli string $S = A_0 A_1 \cdots A_{n-1}$ to be measured,
where $A_i \in \{ \sigma_x, \sigma_y, \sigma_z, I\}$, where $I$ is a two-by-two identity operator.
Conjugating this with a quantum state $|\Psi\rangle = |\psi\rangle|\phi\rangle$,
where, without loss of generality, $|\psi\rangle$ denotes a quantum state of the qubits that are entangled
due to the ansatz and $|\phi\rangle$ denotes a quantum state of the qubits
that are in a classically accessible state, such as a computational basis state,
we may write
\begin{align}
\begin{split}
\label{eqn:qsr}
    \bra{\Psi} S \ket{\Psi} 
    &= \bra{\psi}\bra{\phi} \bigotimes_{i\in P} A_i \bigotimes_{j\in Q} A_j \ket{\psi}\ket{\phi}\\
    & = \bra{\psi} \bigotimes_{i\in P} A_i \ket{\psi}  \bra{\phi} \bigotimes_{j\in Q} A_j \ket{\phi},
\end{split}
\end{align}
where we used sets $P$ and $Q$ to denote 
the set of qubits that are entangled
and the set of qubits that are in a classically accessible state, respectively.
Clearly, the second tensor product in the last line of (\ref{eqn:qsr}) can be
classically efficiently computed. 
Thus, we may evaluate just the first tensor product 
on a quantum computer and this amounts to a reduction
in the number of Pauli strings to measure since there
often are multiple second tensor products with the 
same first tensor product in the full set of
Pauli strings to measure.

As to the latter, we start by a simple example to aid the description. 
Consider a quantum state $|\Psi\rangle$ defined according to
\begin{align}
\label{eqn:bell-state}
    \ket{\Psi} = \sum_{m} \ket{\psi_m}(c_{m,0}\ket{00} + c_{m,1}\ket{11}).    
\end{align}
Since the separated-out, two-qubit states $\ket{00}$ and $\ket{11}$ do not
encode any more than a single qubit of information, this may be compressed to
\begin{align}
\label{eqn:bell-state-compressed}
    \ket{\Psi_{\rm comp}} = \sum_{m} \ket{\psi_m}(c_{m,0}\ket{0} + c_{m,1}\ket{1}).    
\end{align}
Consider now a Pauli string, living in the full space spanned by $\ket{\Psi}$ 
in (\ref{eqn:bell-state}), $S = A_0 A_1 \cdots A_{n-1}$ to measure.
Without loss of generality we designate $A_0 A_1$ as the Pauli product
that lives in the separated-out two-qubit state space in (\ref{eqn:bell-state}).
Conjugating $S$ with $\ket{\Psi}$, we obtain
\begin{align}
\begin{split}
    \bra{\Psi} S \ket{\Psi} 
    &= \sum_{m,n} \bra{\psi_m} \bigotimes_{i>1} A_i \ket{\psi_n} (c^*_{m,0}c_{n,0} \bra{00} A_0 A_1 \ket{00} + c^*_{m,0}c_{n,1} \bra{00} A_0 A_1 \ket{11}  \\
    & \qquad \qquad \qquad \qquad \qquad + c^*_{m,1}c_{n,0} \bra{11} A_0 A_1 \ket{00} + c^*_{m,1}c_{n,1} \bra{11} A_0 A_1 \ket{11}).
\end{split}
\label{eq:Soriginal}
\end{align}
Consider further a compressed Pauli string $S_{\rm comp} = A_{\rm comp} A_2 \cdots A_{n-1}$, 
living in the compressed space spanned by $\ket{\Psi_{\rm comp}}$.
Conjugating $S_{\rm comp}$ with $\ket{\Psi_{\rm comp}}$, we obtain
\begin{align}
\begin{split}
    \bra{\Psi} S_{\rm comp} \ket{\Psi} 
    &= \sum_{m,n} \bra{\psi_m} \bigotimes_{i>1} A_i \ket{\psi_n} (c^*_{m,0}c_{n,0} \bra{0} A_{\rm comp} \ket{0} + c^*_{m,0}c_{n,1} \bra{0} A_{\rm comp} \ket{1} \\
    & \qquad \qquad \qquad \qquad \qquad + c^*_{m,1}c_{n,0} \bra{1} A_{\rm comp} \ket{0} + c^*_{m,1}c_{n,1} \bra{1} A_{\rm comp} \ket{1}).
\end{split}
\label{eq:Scompressed}
\end{align}
Inspecting (\ref{eq:Soriginal}) and (\ref{eq:Scompressed}),
we observe that all that we need to satisfy for the two expressions to agree are
\begin{align}
\bra{00} A_0 A_1 \ket{00} &= \bra{0} A_{\rm comp} \ket{0}, \\
\bra{00} A_0 A_1 \ket{11} &= \bra{0} A_{\rm comp} \ket{1}, \\
\bra{11} A_0 A_1 \ket{00} &= \bra{1} A_{\rm comp} \ket{0}, \\
\bra{11} A_0 A_1 \ket{11} &= \bra{1} A_{\rm comp} \ket{1}.
\end{align}
For all possible $A_0$ and $A_1$, we report in Table~\ref{tab:PauliCompression}
the conversion table for $A_{\rm comp}$. Numerous null matrices that appear
in Table~\ref{tab:PauliCompression}, together with the reduced set of
operators to measure that consists of three single-qubit
Pauli matrices and an identity matrix, significantly reduces
the measurement overhead.

\begin{table}
\centering
\begin{tabular}{ c c c }
\hline \hline
$A_0$     & $A_1$        & $A_{\rm comp}$ \\
\hline
$I$          & $I$          & $I$            \\
$I$          & $\sigma_{x}$ & $0$            \\
$I$          & $\sigma_{y}$ & $0$            \\
$I$          & $\sigma_{z}$ & $\sigma_{z}$   \\
$\sigma_{x}$ & $I$          & $0$            \\
$\sigma_{x}$ & $\sigma_{x}$ & $\sigma_{x}$   \\
$\sigma_{x}$ & $\sigma_{y}$ & $\sigma_{y}$   \\
$\sigma_{x}$ & $\sigma_{z}$ & $0$            \\
$\sigma_{y}$ & $I$          & $0$            \\
$\sigma_{y}$ & $\sigma_{x}$ & $\sigma_{y}$   \\
$\sigma_{y}$ & $\sigma_{y}$ &-$\sigma_{x}$   \\
$\sigma_{y}$ & $\sigma_{z}$ & $0$            \\
$\sigma_{z}$ & $I$          & $\sigma_{z}$   \\
$\sigma_{z}$ & $\sigma_{x}$ & $0$            \\
$\sigma_{z}$ & $\sigma_{y}$ & $0$            \\
$\sigma_{z}$ & $\sigma_{z}$ & $I$            \\
\hline
\end{tabular}
\caption{Conversion table from $A_0 A_1$ to $A_{\rm comp}$ useful for
reducing the number of measurements required for VQE simulations.
$0$ denotes a null matrix.}
\label{tab:PauliCompression}
\end{table}

\section{Number of two-qubit gate distribution}
\label{app:extra-cnot-distribution}

\begin{figure}
\centering
\includegraphics[width=0.95\columnwidth]{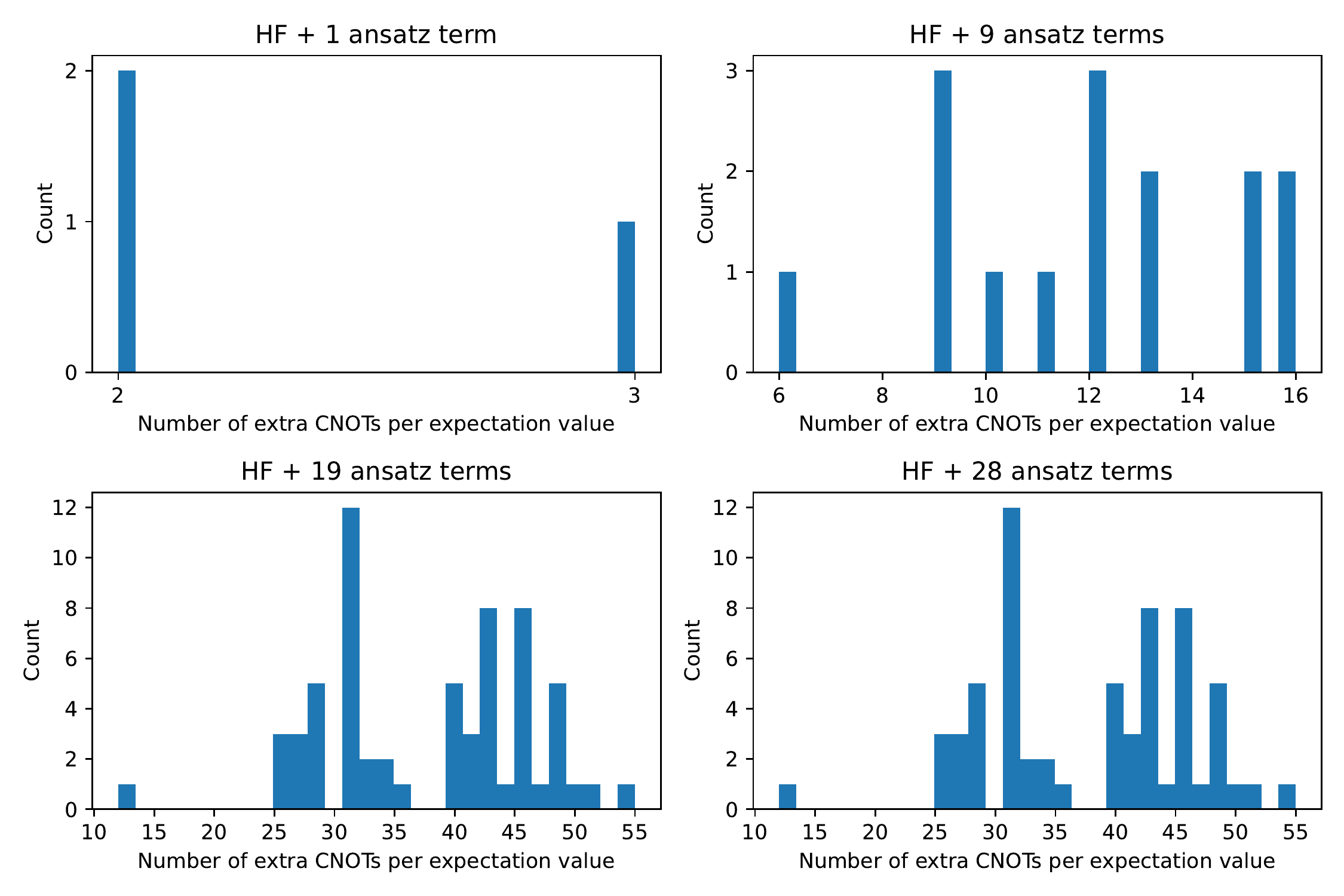}
\caption{Histogram of the extra number of \cnotgates\ required for the GC method
per measurement value.
The Pauli strings to be measured come from the Hamiltonian in HMP2 in (\ref{eqn:perturb-approx}).
Shown are the cases for HF+1, HF+9, HF+19 and HF+28 for the water molecule example
used in the main text.
}
\label{fig:extra-cnot-distribution}
\end{figure}

In this section we briefly discuss the distribution of extra \cnotgates\ required when using GC with QSR(see Sec.~\ref{sssec:prtb_hmp2}).
Note in GC one creates groups of multiple Pauli strings, drawn from the original set of strings required by a VQE simulation, 
that can simultaneously be measured.
Each group can require a different number of extra \cnotgates.
It also requires a different number of measurements, typically smaller
than the number of Pauli strings inside the group.
To illustrate these requirements,
in Fig.~\ref{fig:extra-cnot-distribution} we show the distribution of the number of extra \cnotgates\ 
in the number of required measurements for the GC method for the water molecule example
we use in the main text. We observe that, as expected, the number of extra \cnotgates\
and the number of measurements to perform generally increases as a function of a larger ansatz size.

In the main text Fig.~\ref{fig:perturb-observation}, we opt to use 
the average number of extra \cnotgates\ as opposed to the maximal number
of extra \cnotgates\ for a given ansatz. This is so, since we expect
the quantum computational error expected from each application of
a two-qubit gate to even out between the circuits that require a smaller
than average number of extra \cnotgates\ and the circuits that require a
larger than average number of extra \cnotgates. Thus we select
the average number as a representative number that may best reflect
a practical setting.

\end{document}